\pdfoutput=1
\documentclass{styles/llncs}
\def\modereport{} 

\author{M\'{a}rio S. Alvim\inst{1} \and Piotr Mardziel\inst{2} \and Michael Hicks\inst{3}}
\authorrunning{Alvim, Mardziel, and Hicks}
\tocauthor{M\'{a}rio S. Alvim, Piotr Mardziel, and Michael Hicks}
\institute{Universidade Federal de Minas Gerais, Belo Horizonte, Brazil,
\email{msalvim@dcc.ufmg.br},
\and
Carnegie Mellon University, Pittsburgh, USA,
\email{piotrm@gmail.com}
\and
University of Maryland, College Park, USA,
\email{mwh@cs.umd.edu}
}

\usepackage{amsmath}
\usepackage{amsfonts}
\usepackage{amssymb}
\usepackage{nicefrac}
\usepackage{scalerel}

\usepackage{xcolor}

\usepackage{multirow}

\usepackage[backend=bibtex,maxcitenames=10,maxbibnames=10]{biblatex}

\usepackage{listings}
\usepackage{enumerate}
\usepackage{cancel}
\usepackage{graphicx}
\usepackage[normalem]{ulem}

\usepackage{tikz}
\usepackage{standalone}
\usepackage{import}
\usepackage{setspace}

\usepackage{xspace} 
\usepackage{relsize}

\usepackage[novbox]{pdfsync}

\allowdisplaybreaks 

\definecolor{gray}{rgb}{0.95,0.95,0.95}
\definecolor{darkgray}{rgb}{0.35,0.35,0.35}

\newcommand{\linkform}[1]{\uline{#1}}

\ifdefined\modeprint
 \usepackage[draft=true]{hyperref}
\else
 \usepackage[hidelinks]{hyperref}
 \hypersetup{
      breaklinks=true,
      colorlinks=true,
      linktoc=all,
      linktocpage,
      citecolor=blue,
      filecolor=blue,
      linkcolor=blue,
      urlcolor=blue,
  }
\fi

\makeatletter
\ifHy@draft
\else
\def\hyper@link@[#1]#2#3#4{%
   \protected@edef\Hy@tempa{#2}%
   \ifx\Hy@tempa\@empty
     \hyper@link{#1}{#3}{\linkform{#4}}%
   \else
     \expandafter\hyper@readexternallink#2\\{#1}{#3}{\linkform{#4}}%
   \fi
}

\let\oldfootnotemark\@footnotemark
\def\@footnotemark{\let\inlink y%
\oldfootnotemark%
\let\inlink n%
}
\def\@makefnmark{%
\textsuperscript{
\if\inlink y%
\linkform{\@thefnmark}%
\else%
\@thefnmark%
\fi}}

\fi
\makeatother

\ifdefined\modedraft
  \newcommand{\showindraft}[1]{#1}
\else
  \newcommand{\showindraft}[1]{}
\fi

\newcommand{\msa}[1]{\showindraft{\textcolor{magenta}{MSA -- #1}}}
\newcommand{\mwh}[1]{\showindraft{\textcolor{blue}{MWH -- #1}}}
\newcommand{\pxm}[1]{\showindraft{\textcolor{orange}{PM -- #1}}}

\usepackage{tabulary} 
\usepackage{stackengine} 
\usepackage{subcaption}

\newcommand{\code}[1]{\lstinline!#1!}
\newcommand{\mcode}[1]{\text{\code{#1}}}

\newcommand{\la}{\leftarrow}

\newcommand{\paren}[1]{\left( #1 \right)}

\newcommand{\set}[1]{\left\{ #1 \right\}}

\DeclareMathOperator*{\argmax}{argmax}

\DeclareMathOperator*{\E}{\mathlarger{\mathlarger{\mathbb{E}}}}

\newcommand{\stacklabel}[1]{\stackrel{\smash{\scriptscriptstyle \mathrm{#1}}}}
\newcommand{\defeq}{\stacklabel{def}=}

\newcommand{\dist}[1]{\mathbb{D}#1}

\newcommand{\qm}[1]{``#1''}

\usepackage{mdwlist} 

\usepackage{thmtools}
\usepackage{thm-restate}

\usepackage{wrapfig} 

\usepackage{xifthen} 
\newcommand{\exa}[1][]{\refstepcounter{example}\par\medskip \noindent \emph{Example~\theexample\ifthenelse{\isempty{#1}}{. }{ (#1).}}} 
\newcommand{\exaend}{\medskip} 

\newcommand{\typereal}{by strategy\xspace}
\newcommand{\realsecurity}{security \typereal}

\newcommand{\typeaggr}{by aggregation\xspace}
\newcommand{\aggrsecurity}{security \typeaggr}
\newcommand{\Aggrsecurity}{Security \typeaggr}

\newcommand{\calx}{\mathcal{X}} 
\newcommand{\caly}{\mathcal{Y}} 
\newcommand{\calz}{\mathcal{Z}} 
\newcommand{\cals}{\mathcal{S}} 
\newcommand{\calw}{\mathcal{W}} 
\newcommand{\hyper}[1]{\mathsf{#1}} 

\newcommand{\distsymb}{\mathbb{D}} 

\newcommand{\prior}{\pi} 
\newcommand{\strat}{\pi} 
\newcommand{\m}{\mu} 
\newcommand{\env}{\hyper{En}} 
\newcommand{\model}{\hyper{M}} 
\newcommand{\joint}[1]{#1^{joint}} 
\newcommand{\dprior}{D{\prior}} 

\newcommand{\hyperh}{\hyper{H}} 
\newcommand{\hyperf}{\hyper{F}} 

\newcommand{\distv}{\mathbb{V}} 

\newcommand{\vgf}{V_{g}} 
\newcommand{\bayesvf}{\distv^{(Bayes)}_{X}} 
\newcommand{\contextbayesvf}{\distv^{en(Bayes)}_{X}} 
\newcommand{\strbayesvf}{\distv^{st(Bayes)}_{S}} 
\newcommand{\priorvf}{\distv_{X}} 
\newcommand{\contextvf}{\distv^{en}_{X}} 
\newcommand{\modelvf}{\distv^{md}_{X}} 
\newcommand{\strvf}{\distv^{st}_{S}} 
\newcommand{\strvtempf}{{\strvf}^{(*)}} 
\newcommand{\strvmdlf}{\distv^{st}_{S}} 

\newcommand{\vg}[1]{\vgf(#1)} 
\newcommand{\vgh}[1]{\vgf[#1]} 
\newcommand{\bayesv}[1]{\bayesvf(#1)} 
\newcommand{\contextbayesv}[1]{\contextbayesvf(#1)} 
\newcommand{\strbayesv}[1]{\strbayesvf(#1)} 
\newcommand{\priorv}[1]{\priorvf(#1)} 
\newcommand{\contextv}[1]{\contextvf(#1)} 
\newcommand{\modelv}[2]{\modelvf(#1 , #2)} 
\newcommand{\strv}[1]{\strvf(#1)} 
\newcommand{\strvtemp}[1]{\strvtempf(#1)} 
\newcommand{\strvmdl}[2]{\strvmdlf(#1 , #2)} 

\newcommand{\Avf}{\distv^{(A)}_{X}} 
\newcommand{\contextAvf}{\distv^{en(A)}_{X}} 
\newcommand{\strAvf}{\distv^{st(A)}_{S}} 
\newcommand{\Av}[1]{\Avf(#1)} 
\newcommand{\contextAv}[1]{\contextAvf(#1)} 
\newcommand{\strAv}[1]{\strAvf(#1)} 
\newcommand{\Bvf}{\distv^{(B)}_{X}} 
\newcommand{\contextBvf}{\distv^{en(B)}_{X}} 
\newcommand{\strBvf}{\distv^{st(B)}_{S}} 
\newcommand{\Bv}[1]{\Bvf(#1)} 
\newcommand{\contextBv}[1]{\contextBvf(#1)} 
\newcommand{\strBv}[1]{\strBvf(#1)} 

\newcommand{\abstracts}{\sqsubseteq} 
\newcommand{\compref}{\sqsubseteq_{\circ}} 

\newcommand{\reals}{\mathbb{R}} 

\newcommand{\supp}[1]{\left\lceil #1 \right\rceil} 

\newcommand{\disth}{\mathbb{H}} 

\newcommand{\defeqq}{\stacklabel{def?}=}
\newcommand{\measure}           [0]{information measure\xspace}
\newcommand{\measures}          [0]{information measures\xspace}

\newcommand{\nullmat}{\overline{0}}
\newcommand{\mat}[1]{\left[ #1 \right]}

\newcommand{\environmenth}{\mathbb{H}_{C}}

\ifdefined\modedraft  \fi
\ifdefined\modereport  \fi

\begin{document}     
 
\title{Quantifying vulnerability of secret generation using
  hyper-distributions \ifdefined\modereport (extended version) \fi}
\maketitle              

\begin{abstract}
  Traditional approaches to Quantitative Information Flow \allowbreak (QIF)
  represent the adversary's prior knowledge of possible secret values
  as a single probability distribution.
  This representation may miss important 
  structure.
  For instance, representing prior knowledge about
  passwords of a system's users in this way
  overlooks the fact  that many users generate passwords using 
  some \emph{strategy}.  Knowledge of such strategies can 
  help the adversary in guessing a secret, so ignoring them may
  underestimate the secret's vulnerability.
  In this paper we explicitly model strategies as 
  distributions on secrets, and  generalize the representation of 
  the adversary's prior knowledge from a distribution on secrets
  to an \emph{environment}, which is a distribution on strategies 
  (and, thus, a distribution on distributions on secrets, 
  called a \emph{hyper-distribution}).
  By applying information-theoretic techniques to environments 
  we derive several meaningful generalizations of the traditional
  approach to QIF.
  In particular, we disentangle the 
  \emph{vulnerability of a secret} from the \emph{vulnerability of the strategies} 
  that generate secrets, and thereby distinguish 
  \emph{\aggrsecurity}---which relies on the uncertainty over
  strategies---from \emph{\realsecurity}---which relies on the
  intrinsic uncertainty within a strategy.
  We also demonstrate that, in a precise way, no further generalization
  of prior knowledge 
  (e.g., by using distributions of even higher order) 
  is needed to soundly quantify the vulnerability of the secret.
\end{abstract}


\showindraft{
\newpage
\setcounter{section}{-1} 
\section{Notation and plan of the paper}
\label{sec:outline}

\input{nonfinaltex/sec-outline}
}

\showindraft{
\newpage
}
\section{Introduction}
\label{sec:introduction}


Two core principles within the field of \emph{quantitative information
flow} (QIF) are: (i) a secret is considered \qm{vulnerable} to the extent
the adversary's prior knowledge about secret values has low entropy; and
(ii) the leakage of information in a system is a measure of how much
the observable behavior of the system, while processing a secret value,
degrades that entropy. 
These principles have been used to create ever more sophisticated QIF 
frameworks to model systems and reason about leakage.
(See, for example,
\cite{millen87,mclean90,gray91,clarkhunt01,boreale06,malacaria07,
chatzpalamidessi08a,smith09foundations,koepf11automatically,borealepampaloni11a,Alvim:12:JCS,mciver14abstract,clarkson15integrity}.) 

Traditional approaches to QIF represent the adversary's prior knowledge
as a probability distribution on secret values.
This representation is adequate when secrets are generated according
to a single, possibly randomized, procedure that is known to the 
adversary (e.g., when a cryptographic key is randomly generated according 
to a known algorithm).
However, in some important situations secrets are generated according
to a more complex structure. In these cases, representing the prior as a
distribution loses important, security-relevant information.

Consider the example of passwords. 
If an adversary gains access to a large collection of passwords 
(without the associated user identities), his prior knowledge 
can be modeled as the probability distribution over passwords 
corresponding to the relative frequency of passwords in the collection.
It would be wrong to believe, however, that passwords are
generated by a function exactly described by this distribution.
This representation of prior knowledge aggregates a 
population of users into a single expected probabilistic behavior, 
whereas in fact it is more likely that individual users generate
passwords according to some (not completely random) \emph{strategy}.
Some user born in 1983, 
for instance, may have a strategy of generally 
picking passwords containing the substring \qm{1983}.
If an adversary knows this, he can guess relevant passwords more
quickly. 
In addition, on a system that mandates password 
changes, he may have an advantage when guessing that a changed 
password by the same user contains \qm{1983} as a substring. 
In short, if the adversary learns something about
the secret-generating strategy, he may obtain additional information
about the secret itself.

Generally speaking, knowledge of strategies can be useful 
when multiple secrets are produced by a same source. 
For example, the same user might use a similar strategy to 
generate passwords on different web sites. 
If we consider locations as secret, then
changes in location are surely correlated, e.g., based on time of
day. 
Learning someone's strategy for moving in a city may
increase the chances of guessing this person's location 
at a future point in time.
Perhaps surprisingly, an evolving secret subject to repeated
observations, in some cases, can be learned \emph{faster} if it is
changed (and observed) more often \cite{mardziel14time}. 
The reason is that the strategy by which the secret changes is 
revealed faster if more samples from the strategy are visible to 
an adversary; and if the strategy has little randomness in it, 
the adversary has an increased accuracy in determining past, 
current, and even future secret values.

This paper develops the idea that when secrets are generated according
to a plurality of strategies, as in the above examples, it is
advisable to represent the adversary's prior as a
\emph{hyper-distribution} of secrets, i.e., a distribution of
distributions. To show this, we first define a system model that
explicitly considers strategies for generating secrets.  We formalize
a strategy as a probability distribution from which secrets can be
sampled.  We assume there is a probability distribution on strategies
themselves, which we call an \emph{environment}, representing how
likely it is that each strategy will be used for generating the
secret.  Returning to the password example, each user would have his
own probability distribution for generating secrets (i.e., his own
strategy), and the environment 
would consist in a probability
distribution over these strategies, representing the chance of each
user being the one logging into the system.

In this model, representing the adversary's prior as a distribution on
secrets would reflect the expected behavior of all possible strategies
in the environment. By quantifying the prior vulnerability as a
function of this single distribution, traditional approaches would
miss relevant information, underestimating the vulnerability of the
secret for adversaries able to learn the strategy being used. By modeling
the prior as a hyper-distribution, and applying information-theoretic
reasoning on it, we can do better, generalizing the traditional
approach to QIF\@. 
More specifically, we make the following contributions.

\begin{itemize}

\item We generalize the traditional measure of prior 
adversarial vulnerability to \emph{environmental vulnerability}, 
which takes into account that the adversary can learn
the strategy for generating secrets. (Section~\ref{sec:strategies}.)

\item We define a measure of \emph{strategy vulnerability}, 
which quantifies how certain an adversary is about the 
secret-generating strategy itself. 
%
  We demonstrate that the traditional measure of prior 
vulnerability on secrets neatly decomposes into environmental 
and strategy vulnerability.
Using this decomposition, we are able to disentangle 
two types of security usually conflated in the traditional 
approach to QIF: 
\emph{\realsecurity}, which arises from the intrinsic randomness 
of secret-generating strategies, and \emph{\aggrsecurity}, 
which arises from the adversary's inability to identify particular
strategies in the secret-generation
process. (Section~\ref{sec:security-real-vs-aggregation}.) 

\item We define models of knowledge for adversaries who can 
only partially identify strategies, and we provide measures of 
the vulnerability of the secret and of the strategies themselves 
for this type of adversary. (Section~\ref{sec:models}.)

\item  We demonstrate that the modeling of the adversary's
prior knowledge as a hyper-distribution on secrets is sufficiently
precise: more
complicated models (e.g., distributions on
distributions on distributions on secrets, and such
\qm{higher order distributions}) add no expressive power.
(Section~\ref{sec:hypers-expressiveness}.) 

\item Our work lays a foundation for reasoning about real-world
  scenarios. 
  In this paper we develop an example based on a real password
  dataset. 
  (Section~\ref{sec:example}.)
\end{itemize}

The next section introduces some preliminary concepts while
Sections~\ref{sec:strategies}--\ref{sec:example} present our main
results.
Finally, Section~\ref{sec:related} discusses related work, and
Section~\ref{sec:conclusion} concludes.
\ifdefined\modereport This is an extended version of a conference
paper\cite{alvim17hyper} that includes additional notes about
hyper-distributions and proofs in the appendices. 
\else Full proofs appear in the corresponding technical
report~\cite{alvim17hyperTR}. 
\fi


\showindraft{
\subsection{Piotr's previous modifications on intro}

Two core principles within the field of \emph{quantitative information
  flow} (QIF) are: (i) a secret is considered \qm{safe} to the extent
the probability distribution on secret values has high entropy; and
(ii) the leakage of information in a system is a measure of how much
the observable behavior of the system, while processing a secret
value, degrades the entropy of that secret. 
These principles have been used to create ever more sophisticated QIF
frameworks to model systems and reason about leakage.
However, little attention has been paid to understand the sources that inject
entropy into the distribution on secrets in the first place.
Sources of entropy are often governed by \emph{latent variables}, 
which is information outside of the space of secrets that determine 
how a secret is distributed.  

\begin{example}[Passwords]
  Consider the example of passwords. 
  Users do not generate (completely) random passwords. 
  Rather, a user's password depends to some degree on that user's
  demographics. 
  A person born in 1983, for example, might have a strategy of
  generally picking passwords containing the string \qm{1983}. 
  A person's age is therefore a latent variable. 
  For forthcoming demonstration purposes, let us say that another
  latent variable is a user's gender and that gender does not impact
  at all that user's password. 
\qed
\end{example}

A proper account of the vulnerability of a secret needs to consider
latent variables if they themselves are vulnerable to discovery. 
Suppose an adversary gains access to a large collection of passwords
(without the associated user identities). 
We can derive from such a database a model of adversary prior
knowledge: a probability distribution over passwords. 
This model makes no assumptions about how these passwords were chosen:
effectively, the prior aggregates a population of users into a single
probabilistic behavior. 
But suppose that the adversary knew that passwords with ``1983'' in
them were more likely to have been generated by people born in 1983
and incorporated this latent demographical variable in his model of
passwords. 
This adversary will be advantaged over an adversary which does not
incorporate latent variables in numerous situations:
\begin{itemize}
\item{} If they learned through outside channels the age of a user
  they are attacking.
\item{} If they knew the same user generated passwords for other
  systems (which they attacked).
\item{} If they knew the same user generated multiple passwords for
  the same system over time and they knew something about that user's
  past passwords.
\end{itemize}


The difference between the two adversarial views can be described
using two pairs of pseudo-programs that each generate an identical
distribution of secrets as seen in Figure~\ref{fig:example-passwords}. 
Both (a) in (b) of the figure specify a two-stage process, one for
generating latent variables (which for this example we call
\mcode{birth}), and one for generating secrets called
\mcode{gen_password}. 
In (a), demographical information is latent and is generated at
``birth'' which then determine a distribution of passwords in
\mcode{gen_password}. 
In (b), on the other hand, nothing is latent and the demographics are
presumed to be part of the password generation itself. 
If we presume that \mcode{birth} occurs one for each user, whereas
\mcode{gen_password} occurs many times, the adversary modeling the
world with (a) is advantaged over the one modeling the world with (b)
as (a) could possibly discover latent variables, whereas for (b) there
is nothing to discover: each password in (b) samples a new demographic
independent of past samples.


Latent variables only matter insofar they influence the distribution
of secrets. 
For example, \mcode{gender} above had no impact on the distribution of
secrets in \mcode{gen_password}. 
That is, the distribution \mcode{gen_password(age,male)} is identical
to the distribution \mcode{gen_password(age,female)} for any
\mcode{age}, and learning a user's gender gives an adversary no
advantage.
The various distributions of secrets that arise due to latent
variables we will call \emph{strategies}. 
In the password example, the set of unique strategies in (a) contains
one strategy
\mcode{gen_password(age,male)}=\mcode{gen_password(age,female)} for
each \mcode{age}, and in (b) contains only a single strategy
\mcode{gen_password()}. 
\pxm{Wanted to motivate hypers here but that would require too much
  detail here. 
  Will think about it more.}

The question we are interested in is the following: \textbf{To what
  extent does knowledge of \emph{strategies} help the adversary guess
  the secret?} 
To be able to answer this question, we develop a model that not only
considers the space of possible secrets, but also considers the space
of possible strategies that generated them. 


Generally speaking, knowledge of a secret-generating strategy can be
useful when that strategy is used to produce multiple secrets. 
For example, the same user might use a similar strategy to generate
passwords on different web sites. 
As another example, a new secret may be correlated with an old secret,
according to the strategy used. 
Aside from passwords, if we consider locations as secret, then changes
in location are surely correlated, e.g., based on time of day. 
Perhaps surprisingly, an evolving secret subject to repeated
observations, in some cases, can be learned \emph{faster} if it is
changed (and observed) more often \cite{mardziel14time}. 
The reason is that the strategy by which the secret changes is
revealed faster if more samples from the strategy are visible to an
adversary; and if the strategy has little randomness in it, the
adversary has an increased accuracy in determining past, current, and
even future secret values.

Modeling both secrets and secret-generating strategies lets us
consider adversaries being able to learn the strategy quantify the
advantage this provides them. 
Such an advantage would thus constitute the value in knowing the
strategy as distinct from knowing the secret.

We make the following contributions:
\begin{itemize}
\item{} We show how to model adversary knowledge including latent
  variables as \emph{environments}, composed of strategies within
  hyper-distributions.
\item{} We define vulnerabilities of environments and of strategies
  within them as decomposing prior vulnerability.
\item{} We characterize \realsecurity and \aggrsecurity
  as special cases of the decomposition.
\item{} We define abstraction of environments as sound approximations
  of adversary knowledge and show how to quantify vulnerabilities in
  case of approximate adversary knowledge.
\item{} We demonstrate the concepts using a small numerical examples
  based on a real world password dataset.
\end{itemize}

}

\showindraft{
\newpage
}
\section{Preliminaries}
\label{sec:preliminaries}


%
%
%
%
%
%
%
%
%
%
%
%
%
%

We briefly review standard concepts and notation from
quantitative information flow (QIF). 
Notably we define notions of \qm{secret}, 
an adversary's \qm{prior knowledge} about the secret (or simply, \qm{prior}), 
and an \qm{information measure} to gauge that knowledge. 
We also define \qm{channels}, probabilistic 
mappings from a set of secrets to another set, 
which have the effect of updating the adversary's 
uncertainty about the secret from a prior probability
distribution to a distribution on distributions on secrets, 
called a \qm{hyper-distribution}. 

\subsubsection{Secrets and vulnerability}

A \emph{secret} is some piece of sensitive information we want
to protect, such as a user's password, social security number 
or current location. 
An adversary usually only has partial information about 
the value of a secret, referred to as ``the prior.''
Traditionally, the prior is represented as a probability
distribution; our aim in this paper is to show that an alternative
representation can be more useful.
We denote by $\calx$ the set of possible secrets and by
$\distsymb{\calx}$ the set of probability distributions 
over $\calx$. 
We typically use $\prior$ to denote a probability distribution, and
$\supp{\pi}$ for its support (the set of values with non-zero 
probability).



An \emph{\measure} is a function $\priorvf{:}\distsymb{\calx}{\rightarrow}\reals$ 
mapping distributions on secrets to real numbers.
An \measure can gauge 
\emph{vulnerability}---the higher the value,
the less secure the secret is---or 
\emph{uncertainty}/\emph{entropy}---the higher the value, 
the more secure the secret is. 
There are several definitions of \measures in the literature, varying
according to the operational interpretation of the measure.
Popular instances include
\emph{Bayes vulnerability}~\cite{smith09foundations} and \emph{Bayes risk}~\cite{Chatzikokolakis:08:JCS}, 
\emph{Shannon entropy}~\cite{shannon48communication},
and 
\emph{guessing entropy}~\cite{massey94guessing}.
The \emph{$g$-vulnerability} 
framework~\cite{alvim12gain} was recently introduced to express \measures having richer
operational interpretations; we discuss it further below.

\subsubsection{Hypers and channels}

A \emph{hyper-distribution}~\cite{mciver10compositional} 
(or \emph{hyper} for short) is a distribution on distributions. 
As we will see in the next section, we propose that the prior can be
profitably represented as a hyper.
A hyper on the set $\calx$ is of type $\distsymb^{2}{\calx}$, 
which stands for $\distsymb(\distsymb{\calx})$, a distribution on
distributions on $\calx$.
The elements of $\distsymb{\calx}$ are called the
\emph{inner-distributions} (or \emph{inners}) of the hyper.
The distribution the hyper has on inners is called the
\emph{outer-distribution} (or \emph{outer}).
We usually use $\hyperh$ to denote a hyper,
$\supp{\hyperh}$ for its \emph{support} 
(the set of inners with non-zero probability), 
and $[\prior]$ to denote the point-hyper assigning probability 
$1$ to the inner $\prior$.

An \emph{(information theoretic) channel} is a triple 
$(\calx,\caly,C)$, where $\calx,\caly$ are finite 
sets of input values and output values, resp., 
and $C$ is a $|\calx|{\times}|\caly|$ channel matrix 
in which each entry $C({x,y})$ corresponds to the 
probability of the channel producing output $y$ when 
the input is $x$. 
Hence each row of $C$ is a probability distribution over $\caly$
(entries are non-negative and sum to $1$).
A channel is \emph{deterministic} iff each row contains a 
single $1$ identifying the only possible output for that input.

A distribution $\prior{:}\distsymb{\calx}$ and a channel $C$ 
from $\calx$ to $\caly$ induce a joint distribution
$p(x,y){=}\prior(x)C({x,y})$ on $\calx{\times}\caly$,
producing joint random variables $X, Y$ with marginal 
probabilities $p(x){=}\sum_{y} p(x,y)$ and 
$p(y){=}\sum_{x} p(x,y)$, and conditional probabilities 
$p(y{\mid}x){=}\nicefrac{p(x,y)}{p(x)}$ (if $p(x)$ is non-zero) 
and $p(x{\mid}y){=}\nicefrac{p(x,y)}{p(y)}$ (if $p(y)$ is non-zero). 
Note that $p_{XY}$ is the unique joint distribution that recovers
$\prior$ and $C$, in that $p(x){=}\prior_{x}$ and $p(y{\mid}x){=}C({x,y})$ (if $p(x)$ is non-zero).
\footnote{To avoid ambiguity, we may use subscripts on 
distributions , e.g., $p_{XY}$, $p_{Y}$ or $p_{X \mid Y}$.}
For a given $y$ (s.t. $p(y)$ is non-zero), the conditional 
probabilities $p(x{\mid}y)$ for each $x \in \calx$ form the 
\emph{posterior distribution $p_{X \mid y}$}.

A channel $C$ from a set $\calx$ of secret values 
to set $\caly$ of observable values can be used to model
computations on secrets.
Assuming the adversary has prior knowledge $\prior$ about
the secret value, knows how a channel $C$ works, and
can observe the channel's outputs, the effect of the channel 
is to update the adversary's knowledge from $\prior$ to a 
collection of posteriors $p_{X \mid y}$, each occurring 
with probability $p(y)$. 
Hence, following~\cite{mciver10compositional,mciver14abstract}, we 
view a channel as producing hyper-distribution.%
\footnote{Mappings
of priors to hypers are called \emph{abstract} channels in 
\cite{mciver14abstract}.}
We use $[\pi,C]$ to denote the hyper obtained
by the action of $C$ on $\pi$. 
We say that $[\pi,C]$ is the result of
\emph{pushing prior $\pi$ through channel $C$}.


\subsubsection{Notation on expectations}

We denote the \emph{expected value} of some random
variable $F{:}\calx{\rightarrow}R$  over a distribution 
$\prior{:}\distsymb{\calx}$ by 
$
\E_{\prior}F{\defeq}\E_{x \la \prior}F(x){\defeq}\sum_{x \in \calx} \prior(x)F(x). 
$
Here, $R$ is usually the reals $\reals$ but more generally 
can be a vector space.
If $\calx$ itself is a vector space, then we abbreviate
$\E_{\prior}{(\text{id})}$ by just $\E_{}{\prior}$, the 
\qm{average} of the distribution $\prior$ on $\calx$.

\subsubsection{$g$-vulnerability}

Recently, the \emph{$g$-vulnerability} framework~\cite{alvim12gain}
proposed a family of vulnerability measures that capture
various adversarial models.
Its operational scenario is parameterized by a set $\calw$ of
\emph{guesses} (possibly infinite) that the adversary can make
about the secret, and a \emph{gain function}
$g{:}\calw{\times}\calx{\rightarrow}\reals$. 
The gain $g(w,x)$ expresses the adversary's benefit for having made
the guess $w$ when the actual secret is $x$.
Given a distribution $\prior$, the $g$-vulnerability function measures the
adversary's success as the expected gain of an optimal guessing
strategy:
\begin{align*}
\vg{\prior} 
&\defeq \max_{w \in \calw} \sum_{x \in \calx} \prior(x)g(w,x).
\end{align*}
The $g$-vulnerability of a hyper $\hyperh{:}\distsymb^{2}{\calx}$
is defined as
\begin{align}
\label{eq:vgh}
\vgh{\hyperh}
&\defeq \E_{\hyperh} \vgf.
\end{align}
In particular, when $\hyperh$ is the result of pushing
distribution $\prior{:}\distsymb{\calx}$ through a channel
$C$ from $\calx$ to $\caly$ we have
$
\vgh{\prior,C}
{=}\sum_{y \in \caly} \max_{w \in \calw} \sum_{x \in \calx} \prior(x) C({x,y}) g(w,x).
$

The set of $g$-vulnerabilities coincides with 
the set of all convex and continuous \measures,
which recently have been shown to be precisely those to 
satisfy a set of basic axioms for information measures.%
\footnote{More precisely, if the vulnerability of a hyper 
is defined as the expectation of the vulnerability of its 
inners (as for $\vgf$ in Equation~\eqref{eq:vgh}), 
it respects the data-processing inequality and always 
yields non-negative leakage iff the vulnerability is 
convex.}

\begin{theorem}[Expressiveness of $g$-vulnerabilities \cite{alvim16axioms}]
\label{theo:vggeneral}
Any $g$-vulnerability $\vgf$ is a continuous and
convex function on $\distsymb{\calx}$.
Moreover, given any continuous and convex function 
$\priorvf{:}\distsymb{\calx}{\rightarrow}\reals^{+}$ 
there exists a  gain function 
$g$
with a countable set of guesses 
such that 
$\priorvf{=}\vgf$.
\end{theorem}

%
In the remainder of this paper we will consider only
vulnerabilities that are continuous and convex
(although all of our results carry on for continuous and 
concave uncertainty measures).
We may alternate between the notation $\priorvf$ and $\vgf$ 
for vulnerabilities depending on whether we want to emphasize 
the $g$-function associated with the measure via Theorem~\ref{theo:vggeneral}.

\showindraft{
\newpage
}
\section{Adversarial knowledge as hyper-distributions}
\label{sec:strategies}

This section shows how an adversary's prior knowledge can be
profitably represented as a hyper-distribution on secrets, rather than
simply a distribution. We begin by presenting a basic system model for
wherein secrets are not necessarily generated according to a single
\qm{strategy}, but rather an \qm{environment}, which is a distribution on
strategies. This change motivates an adversary who can learn about the
strategy being used, and from that 
pose a higher threat to the secret.  This notion, which we call
\qm{environmental vulnerability}, strictly generalizes the standard
notion of vulnerability.



\subsection{Strategies and environments}
\label{sec:strategies-environments-contextv}

\begin{wrapfigure}{r}{0.5\linewidth}
\vspace{-9mm}
\centering
\includegraphics[width=0.9\linewidth]{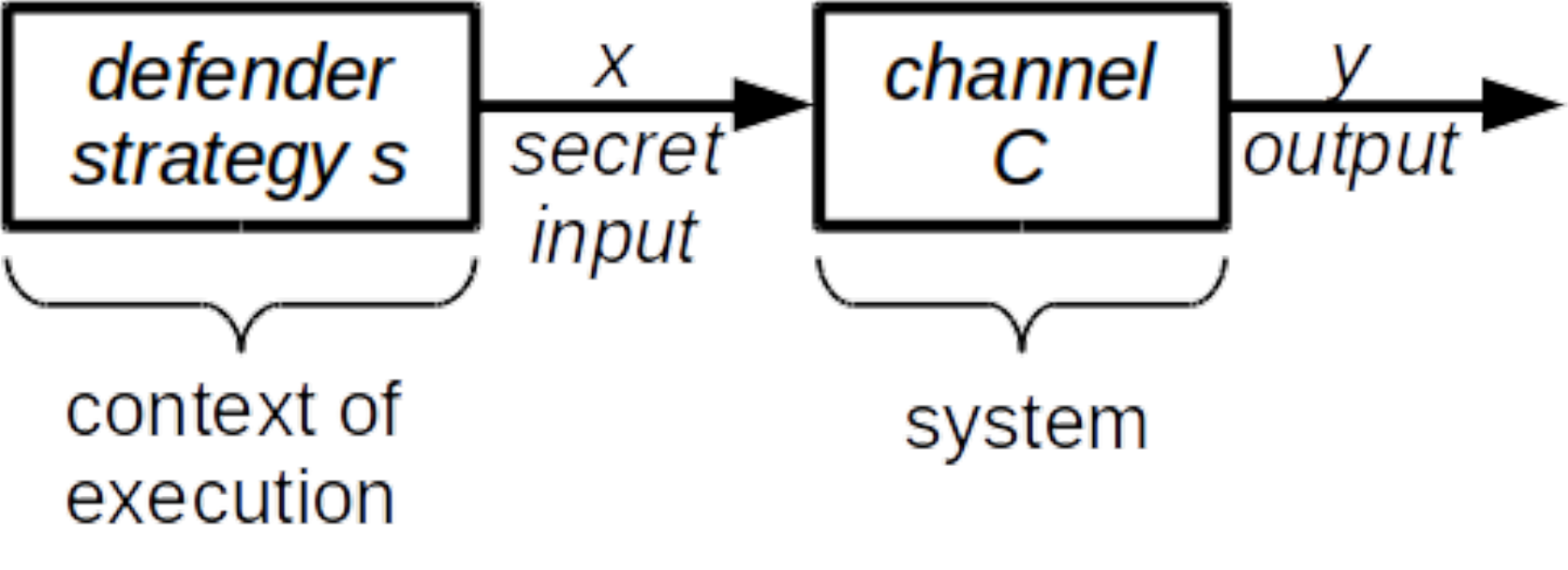}
\vspace{-2mm}
\caption{System and its context.}
\label{fig:context}
\vspace{-8mm}
\end{wrapfigure}
Figure~\ref{fig:context} illustrates our basic model. 
A \emph{system} is a probabilistic mapping from secret inputs to
public outputs, represented as a channel.\footnote{Prior systems often
  also permit public inputs and secret outputs; we leave such
  generalizations to future work.} 
Secrets are produced according to a \emph{strategy} chosen by a
\emph{defender}. 

A strategy is modeled as a probability distribution on the set
of secrets $\calx{=}\set{x_{1}, x_{2}, \ldots, x_{n} }$; i.e., 
the defender chooses the secret by sampling the distribution.
The set $\cals$ of all possible strategies is thus $\distsymb{\calx}$,
but in this paper we shall assume that there is a set
$\cals_{\calx}{=}\{ \strat_{1}, \strat_{2}, \ldots, \strat_{m} \}
\subset \distsymb{\calx}$ of strategies of interest.%
\footnote{Given that $\calx$ is finite, we can make $\cals_{\calx}$
  finite via a discretization that defines an indivisible amount $\mu$
  of probability mass that strategies can allocate among secrets.
  Any precision in strategies can be achieved by making $\mu$ as small
  as needed.}
\mwh{Why are strategies written as $\strat$ here, not $\prior$? 
  Do we need two symbols?}

In traditional QIF, this defender strategy is essentially synonymous
with prior knowledge---we assume the adversary knows exactly
the strategy being used. However, as motivated by the password example
in the introduction, in reality a secret may be generated by a myriad
of possible strategies, and each strategy may be more or less
likely. We represent this idea in our model as an \emph{environment},
which is a probabilistic rule used to   
choose the secret-generating strategy; it
is represented as a probability distribution on the set 
$\cals_{\calx}$ of strategies of interest.
The set $\distsymb{\cals_{\calx}}$ of all possible 
environments is a subset of the set $\distsymb^{2}{\calx}$ 
of all hypers on $\calx$. 
In case 
only one strategy $\prior$ is possible, as in traditional models, the
corresponding environment is the point-hyper $[\prior]$.
We will use letters like $\hyperh$, $\model$, $\env$ to 
denote hypers that are distributions on strategies
of interest.

\begin{wraptable}{r}{0.25\textwidth}
\centering
\vspace{-8mm}
$
\begin{array}{c||c|c|c}
& \strat_1 & \strat_2 & \strat_3 \\ \hline \hline
x_{1}  & 1  & 0               & \nicefrac{1}{2}  \\
x_{2}  & 0  & 1            & \nicefrac{1}{2}  \\ \hline \hline
\env_1 & \nicefrac{1}{2} & \nicefrac{1}{2} & 0 \\ 
\env_2 & 0 & 0               & 1 \\
\end{array}
$
\vspace{-2mm}
\caption{Example~\ref{exa:password1}.}
\label{table:password-1}
\vspace{-8mm}
\end{wraptable}
\exa
\label{exa:password1}
Consider a password-checking system. There are various methods for
choosing passwords, each of which can be represented as a different strategy;
which strategy is used by a particular user is determined
by an environment. 
The adversary is interested in identifying the password used 
for a particular user.
For simplicity, we limit attention to two
possible values for passwords, $\calx{=}\{x_{1}, x_{2}\}$.
Consider the set of possible strategies for generating
secrets is $\cals_{\calx}{=}\{\strat_{1}, \strat_{2}, \strat_{3}\}$, 
where
$\strat_{1}{=}[1, 0]$ always generates secret $x_{1}$,
$\strat_{2}{=}[0, 1]$ always generates secret $x_{2}$, and
$\strat_{3}{=}[\nicefrac{1}{2}, \nicefrac{1}{2}]$ generates either secret
with equal probability.
%
Consider also two possible environments for this system:
\begin{itemize}
\item $\env_{1}{=}[\nicefrac{1}{2}, \, \nicefrac{1}{2}, 0]$ is the environment
in which strategies $\strat_{1}$ and $\strat_{2}$ may be adopted with equal 
probability.
This represents a scenario in which any user logging 
in has an equal probability of having generated his 
password either according to strategy $\strat_{1}$ or 
according to strategy $\strat_{2}$.

\item $\env_{2}{=}[0, \, 0, \, 1]$ is the environment in which strategy 
$\strat_{3}$ is always adopted.
This represents a scenario in which every user logging 
is assured to having generated his password using strategy $\strat_{3}$.
\end{itemize}

We depict strategies and environments in Table~\ref{table:password-1}. 
The columns list strategies; the first grouping of rows contains the
definition of the strategy (i.e., the probability that it chooses a
particular secret), and the next grouping of rows contains the
definition of each environment, one per row, which gives the
probability of each strategy.
\qed
\exaend

\subsection{Prior knowledge as a hyper, and environmental vulnerability}
\label{sec:prior-hyper}


Given a model with an environment $\env$, we can continue to represent
the prior in the traditional manner, as a distribution on secrets
$\prior$. We call this prior the \emph{concise} knowledge of the
environment, and it is defined as the \emph{expectation} of all
strategies of $\env$, i.e., $\prior{=}\E_{}\env$. When this equation
holds, we also say that $\prior$ is \emph{consistent} with $\env$;
when needed we may denote by $\prior_{\env}$ the prior consistent with
environment $\env$.
For instance, consistent, concise knowledge of users' passwords in
Example~\ref{exa:password1} would be the expectation of 
how a randomly picked user would generate their password: 
each user may potentially adopt a unique strategy for 
generating their password, and the prior captures the 
expected behavior of the population of users. 

Alternatively, we can represent the prior as a hyper $\model$,
representing the adversary's \emph{unabridged} knowledge of the
environment $\env$. For now, we will assume an adversary knows the
environment $\env$ precisely, i.e., $\model{=}\env$, just as, in
traditional QIF, it is often assumed that the adversary precisely
knows the defender's single secret-generating strategy. Later, in
Section~\ref{sec:models}, we will introduce the notion of a
\emph{abstraction} $\model$, which is model consistent with an 
environment $\env$, but that does not match it exactly; 
this allows us to model partial adversary knowledge.

Given this new notion of prior (i.e., unabridged knowledge), we must
define a corresponding notion of the vulnerability of a secret. We
call this notion \emph{environmental vulnerability}.


\begin{definition}[Environmental vulnerability]
\label{def:contextv}
  Given a vulnerability measure
  $\priorvf{:}\distsymb{\calx}{\rightarrow}\reals$,
  the \emph{environmental vulnerability} of the secret
  is a function $\contextvf{:}\distsymb^{2}{\calx}{\rightarrow}\reals$ 
  of the environment $\env$ defined as
\begin{align*}
\contextv{\env} ~\defeq~ \E_{\env}{\priorvf}~.
\end{align*}
\end{definition}

It is easy to show that if the environment $\env$ is 
a point-hyper $[\prior]$, environmental vulnerability 
$\contextv{\env}$ collapses into traditional prior vulnerability 
$\priorv{\prior}$.

\begin{restatable}{proposition}{rescollapse}
\label{prop:collapse}
For all environments $\env$, if 
$\env{=}[\prior_{}]$ then
$\contextv{\env}{=}\priorv{\prior_{}}$.
\end{restatable}

The converse of Proposition~\ref{prop:collapse}, however,
is not true, i.e.,
$\contextv{\env}{=}\priorv{\prior}$
does not imply
$\env{=}[\prior]$.
We can also show that, in expectation, an 
adversary with unabridged knowledge $\env$ can never be 
worse-off than an adversary with concise knowledge $\prior_{\env}$.


\begin{restatable}{proposition}{resenvironmentvsdistv}
\label{prop:environmentv-vs-distv}
  For any vulnerability $\priorvf$,
  $\contextv{\env}{\geq}\priorv{\prior_{\env}}$
  for all environments $\env$.
\end{restatable}

Proposition~\ref{prop:environmentv-vs-distv} shows that
the modeling of adversarial knowledge 
as only a distribution on secrets 
overlooks how the adversary can exploit knowledge of the 
environment.
Indeed, as the next example shows, secrets distributed 
according to a same prior may present drastically different 
environmental vulnerability.

\begin{example}
\label{exa:effect-of-context-in-vulnerability}
Consider the password system of Example~\ref{exa:password1}.
Both environments yield the same prior distribution
$\prior 
{=}\E_{}\env_1 
{=}\E_{}\env_2 
{=}[\nicefrac{1}{2},\,\nicefrac{1}{2}]$, 
so an adversary with only concise knowledge would obtain the same
traditional prior vulnerability in both environments.
E.g., for Bayes vulnerability, defined as
\begin{align}
\label{eq:bayesv}
\bayesv{\prior} \defeq \max_{x \in \calx} \prior(x),
\end{align}
the adversary would obtain a traditional prior vulnerability
of $\bayesv{\prior}{=}\nicefrac{1}{2}$.

However, an adversary with unabridged knowledge
would obtain different values for the 
vulnerability of the secret in each environment.
In $\env_1$ environmental vulnerability is
$
\contextbayesv{\env_1}{=}\nicefrac{1}{2}{\cdot}\bayesv{\strat_{1}}{+}\nicefrac{1}{2}{\cdot}\bayesv{\strat_{2}}=\nicefrac{1}{2}{\cdot}1{+} \nicefrac{1}{2}{\cdot}1{=}1
$,
whereas in $\env_2$ environmental vulnerability is
$
\contextbayesv{\env_2}=1{\cdot}\bayesv{\strat_{3}}{=}1{\cdot}\nicefrac{1}{2}{=}\nicefrac{1}{2}
$ (recall that higher is worse for the defender).

Note that in $\env_{2}$, the value for environmental 
vulnerability and traditional prior vulnerability is the same
$
(\contextbayesv{\env_{2}} 
= \bayesv{\prior}{=}\nicefrac{1}{2}),
$
so an adversary who learns the strategy being used is not expected to
be more successful than an adversary who only knows the prior.
%
%
\qed
\end{example}


  


\showindraft{
\newpage
}
\section{\Aggrsecurity and \realsecurity}
\label{sec:security-real-vs-aggregation}

In this section we discuss further the advantage of using a hyper as
the prior, showing how it can distinguish two types of security guarantees 
that are conflated when the prior is merely a distribution:
security \qm{\typeaggr} and security \qm{\typereal}.
We also show that the traditional
definition of prior vulnerability decomposes neatly into 
environmental vulnerability and 
\qm{strategy vulnerability}, which measures the information the
adversary has about the strategy used to generate secrets.

\subsection{Dissecting the security guarantees of traditional prior vulnerability}
\label{sec:motivation-security-real-aggregation}

The final example in the last section provides some insights about
the security guarantees implied by traditional prior vulnerability.
First, \emph{\aggrsecurity} occurs when 
environmental vulnerability (largely) exceeds 
traditional prior vulnerability:
$\contextv{\env}{\gg}\priorv{\prior_{\env}}$.
In this case the secret is protected by the adversary's 
lack of knowledge of the strategy being used, 
and, if the adversary learns the strategy, the vulnerability 
of the secret can (significantly) increase.
An example of \aggrsecurity is a scenario in 
which all users pick passwords with deterministic strategies, 
but the adversary does not know which user is generating the 
password.
If there is a large number of users, and if their strategies 
are varied enough, the passwords may be considered \qm{secure} 
only as long as the adversary cannot use knowledge about the 
environment to identify the strategy being used.

On the other hand, \emph{\realsecurity} occurs when environmental 
and prior vulnerabilities have similar values:
$\contextv{\env}{\approx}\priorv{\prior_{\env}}$.
In this case the secret is protected by the unpredictability 
(or uncertainty) within the strategies that generate the secret,
so even if the strategy becomes known,
the vulnerability of the secret will not increase significantly.
An example of \realsecurity is a bank system in which user PINs are chosen uniformly.
Even if the algorithm is known to the adversary, the vulnerability 
of the secret is not increased.

In Section~\ref{sec:decomposition-vulnerability} we 
define measures of the two types of security discussed above, 
but for that we need first to formalize the concept of 
strategy vulnerability.

\subsection{Strategy vulnerability}
\label{sec:strategy-vulnerability}

We now turn our attention to how the knowledge of an 
environment reflects on the adversary's knowledge 
about the strategy being used to generate secrets. 
For that we will define a measure
$\strvf{:}\distsymb{\cals}{\rightarrow}\reals$ 
of \emph{strategy vulnerability}.

Our measure should cover two key points.
First, it should reflect how certain an adversary 
is about which strategy is being used to generate 
secrets, independently of whether the strategy 
itself is deterministic or random.
In particular, it must distinguish between environments 
in which the adversary knows exactly the strategy being 
used, but that strategy happens to employ randomization 
(in which case strategy vulnerability should be high)
from environments in which the adversary does not know 
what strategy is being used, even if all possible 
strategies are deterministic (in which case strategy
vulnerability should be low).

Second, the measure should characterize environments 
that are \qm{predictable} from the point of view of the
adversary.
The key insight is that $\strv{\env}$ should consider the 
\qm{similarity} among strategies in the support of $\env$.
From the point of view of the adversary, whose goal is to 
\qm{guess the secret} (or, more precisely, to exploit his
knowledge about the secret according to some 
\measure $\priorvf{:}\distsymb{\calx}{\rightarrow}\reals$
of interest), 
two strategies should be considered \qm{similar} if they 
yield \qm{similar} vulnerabilities of the secret, as 
measured according to this $\priorvf$.
The following example motivates this reasoning.

\begin{wraptable}{r}{0.275\linewidth}
\centering
$
\begin{array}{c||c|c|c|c}
& \strat_1 & \strat_2 & \strat_3 & \strat_4 \\ \hline \hline
x_{1}  & 1               & 0 & \nicefrac{1}{2} & \nicefrac{9}{10}\\
x_{2}  & 0               & 1 & \nicefrac{1}{2} & \nicefrac{1}{10} \\ \hline \hline
\env_1 & \nicefrac{1}{2} & \nicefrac{1}{2} & 0               &  0 \\ 
\env_2 & 0 & 0               & 1 & 0 \\
\env_3 & \nicefrac{1}{2} & 0               & 0 & \nicefrac{1}{2} \\
\end{array}
$
\vspace{-2mm}
\caption{Example~\ref{exa:strv-motivation}.}
\label{table:exa-strv-motivation}
\vspace{-8mm}
\end{wraptable}
\exa 
\label{exa:strv-motivation}
Consider an extension from Example~\ref{exa:password1}, 
adding a strategy $\strat_4$ and environment $\env_{3}$,
depicted in Table~\ref{table:exa-strv-motivation}.
Intuitively, strategy vulnerability should be high in
$\env_2{=}[\strat_{3}]$, since an adversary would 
know exactly the strategy being used. 
But what should be the strategy vulnerability in $\env_1$
and in $\env_3$?

Suppose we simply considered the set $\cals_{\calx}$ of 
strategies as our set of secrets, 
and defined $\strvf$ as the Bayes vulnerability w.r.t.
that set:
$
  \strvtemp{\env}{\defeqq}\max_{\strat \in \cals} \env(\strat)~.
$  
As expected we would have $\strvtemp{\env_{2}}{=}1$,
but since in each environment $\env_{1}$ and $\env_{3}$
there are two possible strategies, each with 
probability $\nicefrac{1}{2}$, 
we would then have $\strvtemp{\env_{1}}{=}\nicefrac{1}{2}$, and
$\strvtemp{\env_{3}}{=}\nicefrac{1}{2}$.
But this seems wrong: we are assigning the same measure of 
vulnerability to both $\env_{1}$ and $\env_{3}$, but
these two environments are very different. 
The possible strategies in $\env_{1}$ never
produce the same secret, whereas 
the strategies of $\env_{3}$
produce secrets $x_1$ and $x_2$ with similar
probabilities. $\strvtempf$ ascribes $\env_{1}$ and
$\env_{3}$ the same measure even though the uncertainty
about the strategy under knowledge of $\env_{3}$ seems much lower
than $\env_{1}$. 
For instance, if the adversary is interested in guessing
the secret correctly in one try,
an adversary who knows $\env_{3}$ would always
guess the secret to be $x_1$ and would be right most of the time, 
but an adversary who knows $\env_{1}$ gains no advantage about 
which secret to guess.
In short, for this type of adversary we want
$\strv{\env_{2}}{>}\strv{\env_{3}}{>}\strv{\env_{1}}$, but
$\strvtempf$ fails to satisfy this ordering.
%
\qed
\exaend

These observations lead us to define the vulnerability of 
a strategy in terms of the \emph{difference in accuracy}, 
as measured by a choice of $\priorvf$, of an adversary 
acting according to its full knowledge of the environment 
$\env$ and an adversary betting according to the expected 
behavior $\prior_{\env}{=}\E_{}\env$ of the environment.
The key intuition is that a strategy is, 
\emph{for practical purposes}, known within an environment
when $\priorv{\prior_{\env}}{\approx}\contextv{\env}$, or, 
equivalently,
$\priorv{\E_{}\env}{\approx}\E_{\env} \priorvf$.

\begin{definition}[Strategy vulnerability.]
\label{def:strv}
  Given a vulnerability $\priorvf$,
  the \emph{strategy vulnerability} in environment $\env$
  is defined as the ratio
  \begin{align*}
  \strv{\env} \defeq
  \frac{\priorv{\prior_{\env}}}{\contextv{\env}}. 
  \end{align*}
\end{definition}

By Proposition~\ref{prop:environmentv-vs-distv}, 
$\strv{\env}{\leq}1$, and it is maximum when 
$\priorv{\prior_{\env}}{=}\contextv{\env}$.
As for a lower bound, it can be shown that 
strategy vulnerability is minimum when the adversary's 
measure of interest is Bayes vulnerability.

\begin{restatable}{proposition}{resstrvlowerbound}
\label{prop:strv-lowerbound}
Given any vulnerability $\priorvf$, 
strategy vulnerability is bounded by
$\strv{\env}{\geq}\nicefrac{\bayesv{\prior_{\env}}}{\contextbayesv{\env}}$
for all environments $\env$.
\end{restatable}

\msa{The following example may be shortened, or even go
to an appendix.}

The following example illustrates how Definition~\ref{def:strv} 
covers the two key points.

\begin{example}
\label{exa:strv}
Consider the scenario from Example~\ref{exa:strv-motivation},
but assume an adversary $A$ is only interested in the chances of
correctly guessing the secret in one try, no matter what the 
secret is, whereas an adversary $B$ also wants to guess the 
secret in one try, but considers secret $x_{2}$ as 
$9.5$
times more valuable than secret $x_{1}$ (say, for instance, that 
secrets are passwords to bank accounts, and one of the accounts has 
$9.5$
times more money than the other).

Mathematically, adversary $A$'s measure of success is 
represented by the 
vulnerability $\Avf{=}\bayesvf$ defined in 
Equation~\eqref{eq:bayesv}.
As for adversary $B$, the vulnerability $\Bvf$
can be defined as a $g$-vulnerability where the
set $\calw$ of guesses of guesses is the same as the
set $\calx$ of secrets, and the gain function $g$ is
such that $g(x_{i},x_{j})$ equals $1$ when $i{=}j{=}1$, equals 
$9.5$
when $i{=}j{=}2$, and equals $0$ when $i{\neq}j$.


Table~\ref{table:exa-strv-environments} shows the 
environmental, strategy, and traditional prior vulnerabilities
for each adversary in each environment.
Note that the calculated values substantiate the
intuitions we argued for in Example~\ref{exa:strv-motivation}.
For both adversaries strategy vulnerability is 
maximum in environment $\env_{2}$
($\strAv{\env_{2}}{=}\strBv{\env_{2}}{=}1$), 
and it is higher in environment $\env_{3}$ than 
in environment $\env_{1}$.
\msa{Interestingly, if $x_{2}$ were $20$ times more 
valuable than $x_{1}$, we would obtain 
$\strBv{\env_{3}}{<}\strBv{\env_{1}}$.
This is because in this case the strategy is more
vulnerable in $\env_{1}$ than in $\env_{3}$.
This could be discussed in the journal version.}

In particular for environment $\env_{3}$, the
obtained value $\strAv{\env_{3}}{=}1$ 
meets our intuition that, for practical purposes, adversary 
$A$ has little uncertainty about the strategy being used:
if all he cares about is to guess the secret in one try, 
the differences between the possible strategies are too 
small to provoke any change in $A$'s behavior.
On the other hand, the obtained value 
$\strBv{\env_{3}}{=}\allowbreak\nicefrac{38}{39}{\approx}\allowbreak0.97$ 
reflects our intuition that in the same environment
adversary $B$ has more uncertainty about the strategy being 
used: the differences in each possible strategy 
are significant enough to induce changes in $B$'s behavior.
\begin{table}[tb]
\centering
\begin{small}
$
\begin{array}{c|c||c|c|c||c|c|c}
\multicolumn{2}{c||}{}  & \multicolumn{3}{c||}{\text{Adversary $A$}} & \multicolumn{3}{c}{\text{Adversary $B$}} \\ 
\text{$\env$}  & \text{Prior $\prior_{\env}$}    & \Av{\prior_{\env}} & \contextAv{\env} & \strAv{\env} & \Bv{\prior_{\env}} & \contextBv{\env} & \strBv{\env} \\ \hline \hline
\env_{1} & [\nicefrac{1}{2}, \, \nicefrac{1}{2}]    & \nicefrac{1}{2}   & 1  & \nicefrac{1}{2} & 4\,\nicefrac{3}{4} & 5\,\nicefrac{1}{4} & \nicefrac{95}{105} \\
\env_{2} & [\nicefrac{1}{2}, \, \nicefrac{1}{2}]    & \nicefrac{1}{2}   & \nicefrac{1}{2} & 1 & 4\,\nicefrac{3}{4} & 4\,\nicefrac{3}{4} & 1 \\
\env_{3} & [\nicefrac{19}{20}, \, \nicefrac{1}{20}] & \nicefrac{19}{20} & \nicefrac{19}{20} & 1 & 9\,\nicefrac{1}{2} & \nicefrac{195}{200}  & \nicefrac{38}{39} \\
\end{array}
$
\end{small}
\vspace{2mm}
\caption{Environmental, strategy, and traditional prior vulnerabilities for Ex.~\ref{exa:strv}.}
\label{table:exa-strv-environments}
\vspace{-8mm}
\end{table}
\qed

\end{example}

\subsection{Measures of \aggrsecurity and \typereal}
\label{sec:decomposition-vulnerability}

In this section we provide measures of
the two types of security---\typeaggr and {\typereal}---motivated 
in Section~\ref{sec:motivation-security-real-aggregation}.
The key idea is to observe that Definition~\ref{def:strv}
is consistent with the decomposition of traditional prior vulnerability 
into the product of strategy vulnerability and environmental 
vulnerability, and that these two factors are measures
of \aggrsecurity and \realsecurity,
respectively:
\begin{align}
\label{eq:decomposition-rule-v}
\underbrace{\priorv{\prior}}_{\text{perceived security}} = \underbrace{\strv{\env}}_{\text{\aggrsecurity}} \times \underbrace{\contextv{\env}}_{\text{\realsecurity}}.
\end{align}

Equation~\eqref{eq:decomposition-rule-v} states that any 
fixed amount of traditional prior vulnerability (i.e., \emph{perceived security}) 
can be allocated among strategy and environmental vulnerability 
in different proportions, but in such a way that when 
one increases, the other must decrease to compensate for it.
Environmental vulnerability is a meaningful measure 
of \realsecurity because it quantifies the intrinsic uncertainty
about how secrets are generated within each possible strategy.
Indeed, when strategies are random, this uncertainty
cannot be avoided.
On the other hand, \aggrsecurity 
is a measure of the decrease in the adversary's 
effectiveness caused by his lack of knowledge of the environment.

\begin{example}
\label{exa:decomposition-rule}
Environments $\env_{1}$ and $\env_{2}$ from Example~\ref{exa:strv}
yield the same perceived security for an adversary with
concise knowledge; e.g., for adversary $A$, 
$\Av{\prior_{\env_{1}}}{=}\Av{\env_{2}}{=}\nicefrac{1}{2}$.
However, each environment allocates this perceived security 
differently.
W.r.t. adversary $A$, $\env_{1}$ has minimum \realsecurity 
($\contextAv{\env_{1}}{=}1$), and maximum 
\aggrsecurity ($\strAv{\env_{1}}{=}\nicefrac{1}{2}$).
Conversely, environment $\env_{2}$ has maximum \realsecurity
($\contextAv{\env_{1}}{=}\nicefrac{1}{2}$),
and minimum \aggrsecurity ($\strAv{\env_{1}}{=}1$).
Note that this quantitative analysis precisely characterize
intuitions for the distinction among the two types of security 
motivated in Example~\ref{exa:effect-of-context-in-vulnerability}.
\qed
\end{example}

\mwh{I don't understand why we want the following. If we run out of
  space, perhaps move to the appendix and forward reference this?}
\msa{We have it here because one of the reviewers from FCS, and also
Boris during our presentation, asked whether our definitions are not
just a trivial application of the chain rule for information measures.}  
  
\subsubsection{A note on the chain rule for information measures.}
Equation~\eqref{eq:decomposition-rule-v} 
is not a trivial analogue of the \emph{chain-rule} for information 
measures. 
For a start, most information measures do not follow any traditional
form of the chain rule.%
\footnote{In particular, Bayes vulnerability does not:
in general $V^{(Bayes)}(X,Y){\neq}V^{(Bayes)}(X){\cdot}V^{(Bayes)}(Y \mid X)$.
As an example, consider the joint distribution 
$p$ on $\calx{=}\{x_{1},x_{2}\}$ and $\caly{=}\{y_{1},y_{2}\}$ 
s.t. $p(x_{1},y_{1}){=}\nicefrac{1}{2}$, 
$p(x_{2},y_{1}){=}0$, and
$p(x_{1},y_{2}){=}p(x_{2},y_{2}){=}\nicefrac{1}{4}$.
Then $V^{(Bayes)}(X){=}V^{(Bayes)}(Y\mid X){=}\nicefrac{3}{4}$,
but $V^{(Bayes)}(X,Y){=}\nicefrac{1}{2}$, and the chain rule 
is not respected.}
\msa{Piotr, you were right, what I had here was not the chain
rule, and I fixed the example.}
Even for Shannon entropy, which respects the chain rule,
the decomposition of entropies of random variables $S$, $X$  
corresponding to strategies and secrets, respectively, 
would be
$H(X,S){=}H(S){+}H(X \mid S)$.
But even if it is reasonable to equate $H(X \mid S)$ 
to \qm{environmental entropy} of the secret given the
strategy is known, $H(S)$ cannot be equated with
\qm{strategy entropy} if we want the sum of both
values to be equal to $H(X)$, which is the 
\qm{entropy of the secret}. 
In other words, $H(S)$ does not seem to be a reasonable
measure of \qm{strategy entropy} (in fact, $H(S)$
would be a function on the distribution on strategies
only, so it would fail to take into account the similarity
among strategies).
However, 
we can derive that
$
H(X){=}I(X;S){+}\allowbreak H(X \mid S)
$,
which would suggest that an appropriate measure of
\qm{strategy entropy} is actually $I(X;S)$. 
This is in line with our definition of 
strategy vulnerability as the amount of information
the environment carries about the secret.

\showindraft{
\pxm{adding section to relate to chain rule}

Converting $\priorv{\prior} = \strv{\env} \cdot \contextv{\env}$
into entropy we arrive at:

\begin{equation}
\disth(\prior) = \disth_S{\env} + \environmenth{\env}
\end{equation}

Alternatively, if the entropy measure $ \disth $ respects the chain
rule (min-entropy does not), we can write:
\begin{equation}
\disth(\prior) = \paren{\disth{\env} - \disth(\env | \prior)} + \disth(\prior | \env)
\end{equation}

This gives as two options on how to define $ \disth_S $ and $
\environmenth $:
\begin{itemize}
\item{Option 1}:
  \begin{align*}
    \disth_S{\env} & \defeq -\lg \strv{\env} = -\lg \priorvf(\E{\env}) + \lg \priorvf(\prior | \env)\\
    \environmenth{\env} & \defeq -\lg \contextv{\env} = - \lg \priorvf(\prior | \env)
  \end{align*}
\item{Option 2 (if chain rule applies)}:
  \begin{align*}
    \disth_S{\env} & \defeq \disth{\env} - \disth(\env | \prior) \\
    \environmenth{\env} & \defeq \disth(\env | \prior)
  \end{align*}
\end{itemize}
}

\showindraft{
\newpage
}
\section{Models of adversarial partial knowledge}
\label{sec:models}

Starting from Section~\ref{sec:prior-hyper} we assumed that prior
knowledge represented as a hyper exactly matches the environment
$\env$. However, in real-world settings the adversary is likely only to
know some features of the environment, but not its complete
structure. As such, in this section we develop the notion of a
\qm{model} that is hyper on secrets representing an adversary's
partial knowledge of that environment. 
By employing \qm{abstractions} of the environment as models, we 
are able to generalize prior, environmental, and strategy vulnerability, 
and to provide a stronger version of the \qm{decomposition rule} 
for security of Equation~\eqref{eq:decomposition-rule-v}.

\subsection{Models of partial knowledge as abstractions of the environment}
\label{sec:models-as-abstractions}


A \emph{model of adversarial knowledge} 
is a hyper $\model{:}\distsymb{\cals_{\calx}}$,
representing the adversary's knowledge about how 
secrets are generated.
Each inner $\strat_{j}$ in $\model$ corresponds to
a strategy the adversary can \emph{interpret} as possibly generating a
secret, and the corresponding outer probability 
$\model(\strat_{j})$ represents the probability
the adversary attributes to $\strat_{j}$ being used.

Models can be used to represent states of knowledge
of varied precision.
In particular, the environment $\env$ itself is a model
of an adversary with unabridged knowledge,
whereas the point hyper $[\prior_{\env}]$ 
is the model of an adversary with only concise knowledge.
Here we are interested also in models of intermediate 
levels of adversarial knowledge lying in between 
these two extreme cases. In particular, as we show in the next
example, a model's strategies may not directly match those of the true
environment, but rather abstract information in that environment in a
consistent manner.
\exa
\label{exa:password2}
Consider the password system from Example~\ref{exa:password1}, 
but assume now that the environment $\env$ of execution consists
in six possible strategies, as depicted in 
Table~\ref{table:exa-password-2-env}.
The model of knowledge of an adversary who can always identify 
the user logging into the system is the environment $\env$ itself.
As for an adversary who can never identify the user logging in, 
the model of knowledge is the expected behavior of all users, 
represented by the point hyper $[\prior_{\env}]$ in 
Table~\ref{table:exa-password-2-modelpi}.

\begin{table}[tb]
\centering
\begin{subtable}[b]{0.37\linewidth}
$
\renewcommand{\arraystretch}{1.2}
\begin{array}{c||c|c|c|c|c|c}
& \strat_{1} & \strat_{2} & \strat_{3} & \strat_{4} & \strat_{5} & \strat_{6} \\ \hline \hline
x_{1}  & 1  & 0 & \nicefrac{1}{2} & \nicefrac{1}{4} & \nicefrac{3}{4} & \nicefrac{1}{3} \\
x_{2}  & 0  & 1 & \nicefrac{1}{2} & \nicefrac{3}{4} & \nicefrac{1}{4} & \nicefrac{2}{3}\\ \hline \hline
\env & \nicefrac{1}{10} & \nicefrac{1}{10} & \nicefrac{2}{10} & \nicefrac{3}{10} & \nicefrac{2}{10} & \nicefrac{1}{10} 
\end{array}
\renewcommand{\arraystretch}{1}
$
\caption{Environment $\env$ 
(i.e., model for adversary
with unabridged knowledge).}
\label{table:exa-password-2-env}
\end{subtable}
\hfill
\begin{subtable}[b]{0.275\linewidth}
\centering
$
\renewcommand{\arraystretch}{1.2}
\begin{array}{c||c|c|c}
       & \strat_{A} & \strat_{B} & \strat_{C} \\ \hline \hline
x_{1}  & \nicefrac{1}{2} & \nicefrac{7}{20} & \nicefrac{11}{18} \\
x_{2}  & \nicefrac{1}{2} & \nicefrac{13}{20} & \nicefrac{7}{18} \\ \hline \hline
\hyperf & \nicefrac{2}{10} & \nicefrac{5}{10} & \nicefrac{3}{10}
\end{array}
\renewcommand{\arraystretch}{1}
$
\caption{Model $\hyperf$ for adversary who can identify 
states of the federation.}
\label{table:exa-password-2-modelf}
\end{subtable}
\hfill
\begin{subtable}[b]{0.22\linewidth}
\centering
$
\renewcommand{\arraystretch}{1.2}
\begin{array}{c||c}
       & \prior_{\env} \\ \hline \hline
x_{1}  & \nicefrac{11}{24} \\
x_{2}  & \nicefrac{13}{24} \\ \hline \hline
[\prior_{\env}] & 1
\end{array}
\renewcommand{\arraystretch}{1}
$\caption{Model $[\prior_{\env}]$ for adversary 
with concise knowledge.}
\label{table:exa-password-2-modelpi}
\end{subtable}
\caption{Environment and models of adversary's knowledge for Example~\ref{exa:password2}}
\label{table:exa-password-2-models}
\vspace{-8mm}
\end{table}

Consider now another adversary who cannot exactly identify the user
logging into the system, but can determine from what state in 
the country the user is attempting to login (for instance, by observing the IP of the request).
Assume also that users $\strat_{1}$, $\strat_{2}$ come from 
state $A$, users $\strat_{3}$, $\strat_{4}$ come from state 
$B$, and users $\strat_{5}$, $\strat_{6}$ come from state $C$.
The model of knowledge for this adversary, depicted as 
hyper $\hyperf$ in Table~\ref{table:exa-password-2-modelf},
consists in three strategies $\strat_{A}$, $\strat_{B}$ 
and $\strat_{C}$ representing the expected pattern of password 
generation in states $A$, $B$ and $C$, respectively.
The difference in strategies $\strat_{A}$, $\strat_{B}$ and 
$\strat_{C}$ can capture the different frequency of passwords
from state to state (caused, e.g., by regional
uses of slangs, names of cities, etc.).
The probability assigned by the adversary to each strategy
corresponding to a state is given by the probability of any
given user coming from that state.
For instance, the probability $\hyperf(\strat_{A})$ of 
strategy corresponding to state $A$ is given by
$
\hyperf(\strat_{A})
{=}\env(\strat_{1}){+}\env(\strat_{2})
{=}\nicefrac{1}{10}{+}\nicefrac{1}{10}{=}\nicefrac{2}{10},
$
and strategy $\strat_{A}$ itself is obtained as the 
expectation of all strategies of users coming from 
that state:
$
\strat_{A} 
{=}\nicefrac{\env(\strat_{1})}{\hyperf(\strat_{A})}{\cdot}\strat_{1} 
{+}\nicefrac{\env(\strat_{2})}{\hyperf(\strat_{A})}{\cdot}\strat_{2}
{=}\nicefrac{\nicefrac{1}{10}}{\nicefrac{2}{10}}{\cdot}[1, \, 0] 
{+}\nicefrac{\nicefrac{1}{10}}{\nicefrac{2}{10}}{\cdot}[0, \, 1]
{=}[\nicefrac{1}{2} , \, \nicefrac{1}{2}]
$.
%
%
\qed
\exaend

Model $\hyperf$ of Example~\ref{exa:password2} can
be conveniently represented using a matrix 
representation of hypers as follows.
First, note that any hyper $\hyperh{:}\distsymb{\cals_{\calx}}$ 
induces a joint probability distribution 
$p^{\hyperh}{:}\distsymb{(\calx{\times}\cals_{\calx})}$
on secrets and strategies, defined as 
$p^{\hyperh}(x_{i},\strat_{j}){=}\hyperh(\strat_{j}) \strat_{j}(x_{i})$.
For a hyper $\hyperh$, we let $\joint{\hyperh}$
be the $|\calx|{\times}|\cals_{X}|$ matrix in which
$\joint{\hyperh}({i,j}){=}p^{\hyperh}(i,j)$.
For instance, in Example~\ref{exa:password2}
we have that
\begin{align*}
\joint{\env} = 
\mat{
\begin{array}{cccccc}
\nicefrac{1}{10} & 0                & \nicefrac{1}{10} & \nicefrac{3}{40} & \nicefrac{3}{20} & \nicefrac{1}{30} \\
0                & \nicefrac{1}{10} & \nicefrac{1}{10} & \nicefrac{9}{40} & \nicefrac{1}{20} & \nicefrac{2}{30}
\end{array}
},
\qquad
\text{ and }
\qquad
\joint{\hyperf} =
\mat{
\begin{array}{ccc}
\nicefrac{1}{10} & \nicefrac{7}{40}  & \nicefrac{11}{60}  \\
\nicefrac{1}{10} & \nicefrac{13}{40} & \nicefrac{7}{60}
\end{array}
}.
\end{align*}
Conversely, using the usual concepts of marginalization and
conditioning, given any joint distribution $p^{\hyperh}$ we 
can recover the corresponding hyper $\hyperh$.
Because of that, we shall equate a hyper $\hyperh$
with its corresponding joint distribution $p^{\hyperh}$, 
and, equivalently, with its matrix representation 
$\joint{\hyperh}$.

\begin{wraptable}{r}{0.3\linewidth}
\vspace{-8mm}
\centering
$
A^{State} =
\overbrace{
\mat{
\begin{array}{ccc}
1 & 0 & 0 \\
1 & 0 & 0 \\
0 & 1 & 0 \\
0 & 1 & 0 \\
0 & 0 & 1 \\
0 & 0 & 1 \\
\end{array}
}
}^{\strat_{A} \strat_{B} \strat_{C}}
\left.
\begin{array}{c}
\\
\\
\\
\\
\\
\\
\end{array}
\hspace{-2mm}
\right\}
\begin{array}{c}
\strat_{1} \\
\strat_{2} \\
\strat_{3} \\
\strat_{4} \\
\strat_{5} \\
\strat_{6} \\
\end{array}
$
\vspace{-2mm}
\vspace{-8mm}
\end{wraptable}
Second, 
the adversary's incapability of
distinguishing users within a state can be modeled 
by the matrix $A^{State}$ on the side, which 
maps each strategy corresponding to a user in 
the environment to a strategy corresponding
to a state in the model.
It can be easily verified that the hyper $\hyperf$ in its 
joint form can be recovered as the product of the 
environment $\env$ in its joint form with $A^{State}$, i.e.,
$
\joint{\hyperf} 
{=}\joint{\env}{\times}A^{State}.
$


Although in Example~\ref{exa:password2} the adversary could 
only \emph{deterministically} aggregate strategies together,
in general models can be the result of an adversary
\emph{probabilistically} identifying a trait of the strategy 
used.
Moreover, note that the adversary does not need to know 
the exact strategy from each user first, to only then 
aggregate them into the expected behavior of the state. 
He could, for instance, obtain the average behavior from 
the state directly from a log of passwords in which only 
the user's state of origin is known.

Formally, let $p(\m\mid \strat)$ be the probability of the 
adversary modeling the context as strategy $ \m{:}\cals_{\calx}$ 
when in reality it is strategy $\strat{:}\cals_{\calx}$.
A model $\model$ for environment $\env$ obtained using 
distribution $p(\m\mid \strat)$ assigns to each 
strategy $\m$ outer probability
\begin{align}
\label{eq:abstraction-outer-inner}
\model(\m) = \sum_{\strat} p(\m\mid \strat) \cdot \env(\strat),
\quad
\text{where}
\quad
\m = \sum_{\strat} p(\m\mid \strat) \cdot \strat.
\end{align}


The formulas in Equation~\eqref{eq:abstraction-outer-inner}
are equivalent to the following characterization of the 
abstraction of a model into another in terms of 
\qm{aggregation matrices}.
An \emph{aggregation matrix} $A$ is a
$|\cals_{\calx}|{\times}|\cals_{\calx}|$ channel matrix in which each
entry $A({i,j})$ is the probability $p(\strat \mid  \m)$ of the
adversary mapping strategy $\strat$ to strategy $ \m$.

\begin{definition}[Abstraction of a hyper]
\label{def:abstraction}
A hyper $\hyperh'$ is an \emph{abstraction} of another
hyper $\hyperh$, denoted by $\hyperh'{\abstracts}\hyperh$,
iff 
$
\hyperh'{=}\hyperh{\cdot}A
$
for some \emph{aggregation matrix} $A$. 
\end{definition}

Definition~\ref{def:abstraction} says that an abstraction
$\model$ can be obtained as the result of post-processing 
the environment $\env$ with an aggregation matrix $A$ that 
makes convex combinations of actual strategies.
The matrix $A$ can be seen as the adversary's 
capability of correctly identifying the context of execution.
In particular, when $A$ is the identity matrix $I$, the resulting
abstraction is the environment itself: $\env{=}\env{\cdot}I$.
When $A$ is the non-interferent channel $\nullmat$, the resulting 
abstraction is the point-hyper $[\prior]{=}\env{\cdot}\nullmat$.%
\footnote{The \emph{non-interferent channel} $\nullmat$ is a 
column-matrix in which all rows are identical, and for that 
reason it allows no flow of information from inputs to outputs.
}
In particular, because in Example~\ref{exa:password2} the adversary can
only group whole strategies together based on state, the aggregation
matrix $A^{State}$ is deterministic.

As a sanity check, the following result shows that the 
result of post-processing a hyper with a channel matrix 
is itself a hyper with same expectation, which implies that
all abstractions are consistent with the prior distribution.

\begin{restatable}{proposition}{resaggregation}
\label{prop:aggregation}
If $\hyperh$ is a hyper of type $\distsymb^{2}{\calx}$ and 
$A$ is a channel matrix from $\calx$ to any domain $\caly$, 
then $\hyperh{\cdot}A$ is also a hyper of type $\distsymb^{2}{\calx}$.
Moreover, if we call $\hyperh'{=}\hyperh{\cdot}A$, then
the priors from both hypers are the same: 
$\prior_{\hyperh}{=}\prior_{\hyperh'}$.
\end{restatable}

%
%

\subsection{Vulnerability of the secret given an abstraction}
\label{sec:distv-given-model}


We will now generalize the definition of 
environmental vulnerability of the secret (in which
the adversary is assumed to have unabridged knowledge),
to scenarios in which the adversary's knowledge is an 
abstraction $\model$ of the environment $\env$.

The key insight of this measure is that, whereas
the adversary's actions are chosen depending on his 
modeling of the context as strategy $ \m$ from $\model$, 
his actual gain should be measured according to the real 
strategy $\strat$ coming from the environment $\env$.
We formalize this below, recalling that, from 
Theorem~\ref{theo:vggeneral} we know that 
every continuous and convex vulnerability $\priorvf$
can be written as a $g$-vulnerability $\vgf$ for some
suitable $g$.

\begin{definition}
  \label{def:vul-model-env}
  \emph{The vulnerability of the secret in an environment $\env$ when
    the adversary's model is abstraction $\model$} is given by
  \begin{align}
    \label{eq:modelv}
    \modelv{\model}{\env}
      &= \sum_{\strat} \env(\strat) \sum _{ \m} A({ \m,\strat}) \sum_{x} \strat({x}) \, g(w_{ \m},x), 
  \end{align}
  where
  $ w_{\m}{=}\argmax_{w} \sum_{x} \m(x) g(w,x) $ is the
  adversary's optimal guess if the secret were actually distributed
  according to strategy $ \m$.
\end{definition}

Note that Equation~\eqref{eq:modelv} is defined
only when $p(\m\mid \strat){=}A({\m,\strat})$ is well 
defined, that is, when there exists an aggregation
matrix $A$ making $\model{\abstracts}\env$.

The following result states that the vulnerability
of the secret for an adversary who reasons
according to an abstraction (as per Equation~\eqref{eq:modelv})
is the same as environmental vulnerability
in case this abstraction were the real environment.

\begin{restatable}{proposition}{resenvironmentgen}
\label{prop:environmentv-gen}
For any vulnerability $\priorvf$, environment $\env$
and model $\model$, if $\model{\abstracts}\env$ then
$\modelv{\model}{\env}{=}\contextv{\model}$.
\end{restatable}

\pxm{Can we get an example of the inverse of that, so if not
  $ \model \cancel{\abstracts}\env $, then the equation might not
  hold?
  This would demonstrate the usefulness of the abstractions for
  modeling adversary knowledge.}
\msa{If $ \model \cancel{\abstracts}\env $ then, by definition there
  is no valid conditional distribution $p( \m\mid \strat)$ for all
  strategies $\m\in \model$ and $\strat \in \env$.
  That means that the quantity $\modelv{\model}{\env}$ is not well
  defined, because it needs the distribution $p(\m\mid \strat)$.
  To demonstrate what you are suggesting, I think that we first need
  an approximation of $\modelv{\model}{\env}$ when $p(\m\mid \strat)$
  is not defined, and only then we can investigate how it relates to
  $\contextv{\model}$.
  Here there is a rough idea of how to find such an approximation,
  which is based on the intuition that if the adversary's model
  $\model$ is not an abstraction for $\env$, then $\model$ split up
  strategies that should not have been split up, and we can try to
  find an abstraction $\model'$ for $\env$ which is obtained with the
  \qm{minimum amount of work} to \qm{undo} all inconsistent
  splittings.
  If $ \model \cancel{\abstracts}\env $ then we know that every matrix
  $A$ such that $\env \cdot A{=}\model$ is not a channel matrix, that
  is, it contains values $A({\strat, \m})$ outside the range $[0,1]$,
  which means that $p(\m\mid \strat)$ is not well defined for these
  values.
  We could identify all columns $ \m$ in $A$ for which
  $p(\m\mid \strat)$ is not well defined for some $\strat$ as evidence
  that these $ \m$'s are inconsistent with the environment. 
  We can then group all these invalid $ \m$'s into an aggregated new
  strategy $ \m'$ (and I believe we can easily show that by doing that
  $p( \m')$ and $ \m'$ will be well defined).
  By replacing all inconsistent $ \m$'s with an aggregated $\m'$ we
  obtain a valid aggregation matrix $A'$ for which the alternative
  model $ \model'{=}\env \cdot A'$ is a valid abstraction.
  We can then measure $\modelv{\model}{\env}$ in terms of
  $\modelv{\model'}{\env}$.
  The trick here is that it seems there are several ways to obtain
  $A'$ from $A$ (and there may be several $A$'s too), so we should be
  cautious and choose among all options the most refined $A'$ to make
  sure we are choosing $\model'$ as the most refined abstraction
  compatible with $\model$.
  In other words, we need to make sure that $\model'$ represents the
  most advantageous abstraction for the adversary, so we do not
  underestimate the maximum vulnerability the adversary can obtain
  with the knowledge he has.}

Proposition~\ref{prop:environmentv-gen} has a few interesting
consequences. 
\msa{
First, it implies that an adversary reasoning according
to abstraction $\model$ can precisely compute the
vulnerability of the secret given his model, even without
knowing what the environment is.
}
First, it implies that 
the definition of $\modelv{\model}{\env}$ 
generalizes environmental and traditional prior vulnerabilities:
when the adversary's model is $\model{=}\env$, 
we have that $\modelv{\model}{\env}{=}\allowbreak\contextv{\env}$,
and his model is $\model{=}[\prior_{\env}]$, 
we have that
$\modelv{[\prior]}{\env}{=}\contextv{[\prior]}{=}\priorv{\prior}$.

More importantly, though, 
Proposition~\ref{prop:environmentv-gen} provides a precise information-theoretic characterization 
of our definition of abstractions for an environment.
More precisely, it can be used to show that 
by using a more refined model an adversary can never
be worse off than by using a less refined model.

\begin{restatable}{proposition}{resordermodels}
\label{cor:order-models}
If $\model', \model$ are abstractions for an environment
$\env$, then 
$\model'{\abstracts}\model$ iff
$\modelv{\model'}{\env} {\leq}\modelv{\model}{\env}$
for all vulnerabilities $\priorvf$.
\end{restatable}


\subsection{Strategy vulnerability given an abstraction}
\label{sec:strv-given-model}

Next, we will generalize strategy vulnerability
to the scenario in which the adversary reasons according
to an abstraction $\model$ of the environment $\env$.

Our definition is analogous to that of strategy 
vulnerability, and it is based on the observation that
a strategy is vulnerable given a model to the extent 
the average behavior of the model can be used to infer 
the strategy being used.
In other words, the strategy is protected if
knowledge about the model does not give 
information about what strategy is being used.

\begin{definition}
\label{def:strv-given-model}
Given a vulnerability $\priorvf$, the corresponding 
\emph{strategy vulnerability given an abstraction} 
$\model$ within an environment $\env$ is 
defined as
\begin{align*}
\strvmdl{\env}{\model}
&\defeq \frac{\modelv{\model}{\env}}{\contextv{\env}} 
= \frac{\contextv{\model}}{\contextv{\env}},
\end{align*}
where the second equality stems from Proposition~\ref{prop:environmentv-gen}.
\end{definition}

The next result shows that a more refined 
abstraction never yields smaller strategy vulnerability 
than a less refined abstraction for the same 
environment.

\begin{restatable}{proposition}{resstrategyvulnraiblityrefinement}
\label{prop:strategy-vulnerability-refinement}
Given two abstractions $\model$ and $\model'$ of an
environment $\env$,
$\model'{\abstracts}\model$ iff 
$\strvmdl{\model'}{\env}{\leq} \strvmdl{\model}{\env}$
for all vulnerabilities $\priorvf$.
\end{restatable}

Proposition~\ref{prop:strategy-vulnerability-refinement}
implies bounds on strategy vulnerability given an abstraction.

\begin{restatable}{proposition}{resboundsstrvgivenmodel}
\label{prop:bounds-strategy-vulnerability-given-model}
Given any vulnerability $\priorvf$, for any environment $\env$ 
and any abstraction $\model{\abstracts}\env$,
$\strv{\env}{\leq}\strvmdl{\model}{\env}{\leq}1$,
with equality for the lower bound occurring when $\model{=}[\prior_{\env}]$,
and equality for the upper bound occurring when $\model{=}\env$.
\end{restatable}

Finally, we note that Definition~\ref{def:strv-given-model}
naturally extends the decomposition rule of 
Equation~\eqref{eq:decomposition-rule-v} and the definitions
of different types of security as follows.
\begin{align*}
\underbrace{\modelv{\model}{\env}}_{\text{\stackanchor{perceived security}{given a model}}} 
&= \underbrace{\strvmdl{\env}{\model}}_{\text{\stackanchor{\aggrsecurity}{given a model}}} 
\times \underbrace{\contextv{\env}}_{\text{\stackanchor{\realsecurity}{given a model}}}.
\end{align*}


\msa{The following notes could be cut is space is needed.}
\subsubsection{An interesting observation.}
\label{}

The following observation means that the increase
in accuracy given by a more refined abstraction 
$\model$ over a less refined abstraction $\model'$
is the same for secrets and for strategies.
If $\model'{\abstracts}\model{\abstracts}\env$ then
\begin{align}
\label{eq:ratio}
\frac{\strvmdl{\env}{\model'}}{\strvmdl{\env}{\model}}
&= \frac{\contextv{\model'}}{\contextv{\env}} \times \frac{\contextv{\env}}{\contextv{\model}} 
= \frac{\contextv{\model'}}{\contextv{\model}}.
\end{align}

Making $\model{=}\env$ in Equation~\eqref{eq:ratio}
we recover the definition of strategy vulnerability:
$
\strv{\env}
{=}\nicefrac{\priorv{X}}{\contextv{\env}}
$.
Making $\model'{=}[\prior_{\env}]$ in Equation~\eqref{eq:ratio}
we obtain that the increase in information about secrets
and the increase in information about strategies provided
by a model is the same:
$
\nicefrac{\strv{\env}}{\strvmdl{\env}{\model}}
{=}\nicefrac{\priorv{X}}{\contextv{\model}}
$.

\msa{It'd be nice to have an example in which
$\strv{\env} > \priorv{X}$ and another in which
$\strv{\env} < \priorv{X}$, but in both examples
the increase in information is the same about $\env$
and about $X$.
\ \\
\begin{example}
In environment $\env$ below 
$$
\begin{array}{c||c|c}
       & \strat_1 & \strat_2   \\ \hline \hline
x_{1}  & 0.54 & 0.48 \\
x_{2}  & 0.46 & 0.52 \\ \hline \hline
\env & \nicefrac{1}{2} & \nicefrac{1}{2} 
\end{array}
$$
if we pick $\priorvf$ to be Bayes vulnerability
then $\strv{\env}{>}\priorv{X}$, but if we pick
$\priorvf$ to be $V_g$ for $g$ being the double of
$g_{id}$, then $\strv{\env}{<}\priorv{X}$.
\ \\
\qed
\end{example}
}


\showindraft{
\newpage
}
\section{On the expressiveness of hypers}
\label{sec:hypers-expressiveness}


\newcommand{\hyp}{\pi}
\newcommand{\hypn}[1]{\hyp^{#1}}
\newcommand{\hypnvf}[1]{\distv^{#1}}
\newcommand{\hypnv}[2]{\hypnvf{#1}(#2)}
\newcommand{\hypstar}{\widehat{\hyp}^{2}}
\newcommand{\jointh}[2]{p_{#2}}
\newcommand{\collapseh}[1]{\widehat{\hyp}^{#1}}

\pxm{I think this section can be reduced to the next few paragraphs
  and moved to somewhere earlier in the paper, no need to make it a
  whole section:}

\showindraft{Hyper distributions play a big part in this paper to
  model distributions over strategies. 
  Having gone from distributions over secrets to distributions over
  distributions over secrets, one might wonder whether further levels
  of distribution might be necessary to fully account for adversary
  knowledge. 
  The simple answer is no.

  First note that a hyper-distribution $ \dist\paren{\dist{X}} $ over
  a set $ X $ is a reduced representation of a joint distribution
  $ \dist\paren{X \times Y} $ for some set $ Y $. 
  The inners in a hyper are the unique set of conditioned
  distributions $ p(\cdot | y) $ and the outer probability of an inner
  $ p(\cdot | y) $ is the sum total of all $ p(y') $ where
  $ p(\cdot | y') = p(\cdot | y) $. 
  An alternate formulation of the models in this paper could use joint
  distributions $\dist\paren{X \times Y}$ instead, where $ X $ are the
  secrets and $ Y $ is some latent variable correlated with $ X $ (for
  example a users age in the password example). 
  Environmental vulnerability could then be defined by conditioning on
  $ Y $ instead of using an expectation over inners. 

  In a similar manner, we could model settings with multiple latent
  variables using distributions
  $ \dist\paren{X \times Y_1 \times Y_2 \times \cdots \times Y_n} $ or
  using hyper-hyper-distributions $ \dist^{n+1}{X} $. 
  Our definitions, however, make only a distinction between secrets
  and everything else; they do not make a distinction between
  different latent variables.
  A distribution
  $ \dist\paren{X \times Y_1 \times Y_2 \times \cdots \times Y_n} $
  can just as well be seen as a distribution
  $ \dist\paren{X \times Y}$ with a single latent variable
  $ Y = Y_1 \times Y_2 \times \cdots \times Y_n$, or a hyper
  $ \dist^2 X $.

}
\pxm{the original}

Hyper distributions play an essential role in this paper to
generalize the modeling of secret-generation process 
and the adversary's prior knowledge about it.
Having gone from distributions over secrets to distributions over
distributions over secrets, one might wonder whether further levels
of distribution (i.e., \qm{higher-order} hypers of type 
$\distsymb^{n}{\calx}$, for $n{>}2$) 
might be necessary to fully account for adversary knowledge. 
The simple answer is no.

The core idea is that a hyper corresponds to a 
joint distribution in $\dist(\calx{\times}\caly)$ for some set 
$\caly$ of labels for distributions on $\calx$.
Likewise, an object of type $\distsymb^{n+1}{\calx}$ corresponds 
to a joint distribution in
$\distsymb(\calx{\times}\caly_1{\times}\cdots{\times}\caly_{n})$,
which is itself equivalent to a joint distribution in
$\distsymb(\calx{\times}\caly)$ where
$\caly{=}\caly_1{\times}\cdots{\times}\caly_{n}$.
But note that $\distsymb(\calx{\times}\caly)$ is 
equivalent to a hyper of type $\distsymb^2{\calx}$.
Hence, any 	\qm{higher-order} hyper is equivalent to some 
regular hyper of type $\distsymb^{2}{\calx}$ and,
moreover, both objects preserve the same distribution
on distributions on $\calx$.
Since measures of the vulnerability of the secret are 
functions of distributions on $\calx$, the user of 
\qm{higher-order} hypers is not necessary to 
measure vulnerability.

To make this idea precise, let $\hypn{n}$ range
over objects of type $\distsymb^{n}\calx$.
If the adversary's knowledge is represented by $\hypn{n}$ 
(for some $n{\geq}2$), it is natural to define 
the vulnerability of the secret as the expectation of the 
vulnerabilities of hypers of lower order.
A \emph{vulnerability of order $n$} is a function 
$\hypnvf{n}{:}\distsymb^{n}\calx{\rightarrow}\reals$ 
s.t. $\hypnv{1}{\hypn{1}}{=}\priorv{\hypn{1}}$, and
$\hypnv{n}{\hypn{n}}{=}\E_{\hypn{n}}\hypnvf{n-1}$ for $n \geq 2$.
In particular, $\hypnv{1}{\hypn{1}}{=}\allowbreak\priorv{\hypn{1}}$ is the 
traditional vulnerability on secrets, 
and $\hypnv{2}{\hypn{2}}{=}\E_{\hypn{2}}\priorvf$ 
is environmental vulnerability.
The next result shows that an adversary who reasons according 
to a model of type $\distsymb^{n}{\calx}$ for some $n{\geq}2$
is only as well off as an adversary with an appropriate
model of type $\distsymb^{2}{\calx}$.

\begin{restatable}{proposition}{reshypersexpressivenes}
\label{prop:hypers-expressiveness}
For every $\hypn{n}{:}\distsymb^{n}{\calx}$,
with $n{\geq}2$, 
$\hypnv{n}{\hypn{n}}{=}\hypnv{2}{\collapseh{2}}$,
where $\collapseh{2}{:}\distsymb^{2}{\calx}$ is the hyper 
resulting from marginalizing the joint of $\hypn{n}$ w.r.t.
$Y_{2}{\times}Y_{3}{\times}\ldots{\times}Y_{n-1}$.
\end{restatable}

\msa{Piotr's original notes are not in
  Appendix~\ref{sec:hypers-justification-piotr}.}


\showindraft{
\newpage
}
\section{Case study}
\label{sec:example}


To illustrate the utility of our model, we synthesize an
environment based on the RockYou password dataset~\cite{rockyou}, 
which contains the un-hashed passwords of around 32 million 
users of the RockYou gaming site. 
We construct several abstractions for this environment,
computing for each of them the corresponding vulnerability
of the secret and strategy vulnerability, and show
how they relate.

To synthesize the environment, we begin by reducing the 32 
million passwords to the around 350 thousand passwords that 
contain a string suggesting the birth year of the password's 
owner (the strings \qm{1917} through \qm{1995}). 
We assume that each of these passwords was generated by a 
distinct user, and construct a deterministic strategy for 
each of these users.
The intention is that each strategy represents the user's 
exact preference at the time they selected their password.
The environment consists in these strategies distributed
according to their relative frequency in the database.

To construct abstractions for this environment, we attribute 
to each user the birth year used in their password, as well as 
a randomly chosen gender.
The first abstraction, called \textbf{Omniscient}, is the 
environment itself, and it represents an adversary with 
unabridged knowledge.
Although this level of knowledge is beyond any 
realistic adversary, it will illustrate the limiting values 
of vulnerability.

To construct the \textbf{Age} abstraction, we partition users 
into blocks according to their birth year.
From each block we derive a distribution on passwords representing 
the expected strategy for a person born in that year. 
This produces one strategy for each birth year from 1917 through 1995,
and the probability of each strategy is determined by the relative
frequency of each birth year.

The \textbf{Gender} abstraction aggregates users by gender,
and contains one strategy representing the expected behavior 
of males and of females.
Since we assigned genders to users uniformly at random, these 
two strategies each occur with equal probability ($0.5$) and are 
mostly similar.

Finally, the \textbf{Prior} abstraction has only one strategy 
in its support that aggregates all of the 350 thousand users, 
with each password's probability being proportional to its 
relative frequency. 
This environment is equivalent to the point hyper $[\prior]$ 
containing only the prior distribution on secrets.

\begin{wrapfigure}{r}{0.6\linewidth}
\vspace{-8mm}
\centering
\includegraphics[width=\linewidth]{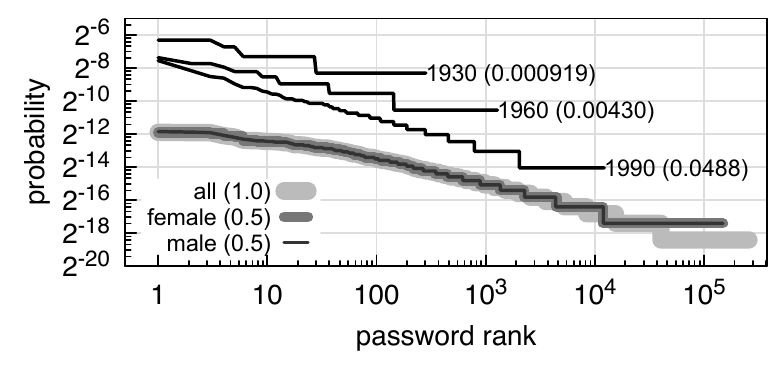}
\vspace{-8mm}
\caption{Example strategies and their probabilities in several
  environments.
  }
\label{fig:environments}
\vspace{-8mm}
\end{wrapfigure}
Several strategies in the last three abstractions are visualized 
in Figure~\ref{fig:environments}. 
The \qm{all} line shows the probability of various passwords being
picked in the \textbf{Prior} environment, sorted by their rank 
(most probable first). 
The two gender aggregate strategies from the \textbf{Gender}
environment are labeled \qm{male} and \qm{female} (note that 
\qm{male}, \qm{female} and \qm{all} largely coincide). 
Finally, three example years from the \textbf{Age} environment are
labeled \qm{1930},  \qm{1960}, and  \qm{1990}. 
The Bayes vulnerability of each strategy is the probability of the
rank 1 password and min-entropy is negation of the base 2 exponent of
that probability.

\begin{table}[tb]
\centering
\begin{tabular}{r|l@{~~}c@{~~}l@{~~}c@{~~}l}
           & $ \bayesv{\prior} $ & $ = $ & $ \contextbayesv{\env} $ & $ \times $ & $ \strbayesv{\env}$  \\[1ex] \hline \\[-1ex]
Omni       & $ 2^{-11.892}   $ & $ = $ & $ 2^{-0}      $ & $ \times $ & $ 2^{-11.892} $ \\
Age        & $ 2^{-11.892}   $ & $ = $ & $ 2^{-7.442}  $ & $ \times $ & $ 2^{-4.450}  $ \\
Gender     & $ 2^{-11.892}   $ & $ = $ & $ 2^{-11.876} $ & $ \times $ & $ 2^{-0.0158} $ \\
Prior      & $ 2^{-11.892}   $ & $ = $ & $ 2^{-11.892} $ & $ \times $ & $ 2^{-0}      $ \\
\end{tabular}
\caption{Bayes vulnerability decomposition.}
\label{table:vul-decomposition}
\vspace{-10mm}
\end{table}
The decomposition of prior Bayes vulnerability as per
Definition~\ref{def:strv} is summarized in
Table~\ref{table:vul-decomposition}. 
Note that the vulnerability in the prior is around
$2^{-11.892}=2.632{\cdot}10^{-4}$.
An adversary who can learn the user's gender could achieve
vulnerability of $2^{-11.876}{=}2.66084\cdot 10^{-4}$.
The strategy vulnerability here shows negligible advantage over the
prior as we synthesized the gender uniformly.
On the other hand, an adversary reasoning according to the aggregation
by age, the vulnerability of the secret is
$ 2^{-7.442}{=}57.526 \cdot 10^{-4} $, providing the equivalent of
$4.450$ bits of information over the prior when measured as
min-entropy.

These measurements let us reach several conclusions. 
First, the (environmental) vulnerability of the prior forms a baseline
level of security in the authentication system for the users in this
experiment. 
The measurements for age and gender abstractions, on the other hand,
gauge the effective security under the pessimistic assumption that
users' age or gender (respectively) can be discovered by an adversary.
The complement (strategy vulnerability) of these measurements 
give the relative importance of keeping these demographics secret. 
In this case, gender is unimportant, while age encodes a significant
amount of a password's entropy. 
A system designer should be wary of displaying age on user profiles.

%
%




\showindraft{
\newpage
}
\section{Related work}
\label{sec:related}


Our work is mainly motivated by the questions raised by the model of
Mardziel et al.~\cite{mardziel14time} for dynamic secrets that evolve
over time, and that may vary as the system interacts with its
environment.
Their model also considers secrets that are generated according to a
strategy, and they give an example that an evolving secret subject to
repeated observations, in some cases, can be learned faster if it is
changed (and observed) more often.
They suggest that this effect is related to the lack of randomness
within the strategy for generating secrets, but they do not develop a
formal measure of that randomness.
In~\cite{mardziel14loss} the authors take a step further and
distinguish between adversary's and defender's goals,
but they still do not have results about the vulnerability of the strategy itself.

Hyper-distributions were introduced in~\cite{mciver10compositional} to
model the adversary's posterior knowledge about the secret (i.e.,
after an observation of the system is performed).
The inners of the hyper are conditional distributions on secrets given
each possible observable produced by the system, and the outer is a
distribution on the observables.
Several other models for QIF have used hypers in a similar way
(e.g.,~\cite{Alvim:14:CSF,mciver14abstract,alvim16axioms}), but all of
them still model prior knowledge as a single distribution on secrets. 
Our work models prior knowledge itself as a hyper-distribution, in
which the inners are strategies for generating secrets, and the outer
is a distribution on strategies.

Several models investigate systems in which secrets are correlated in
interactive systems.
Some approaches capture interactivity in systems by encoding it as a
single \qm{batch job} execution. 
Desharnais et al.~\cite{Desharnais:02:LICS}, for instance, model the
system as a channel matrix of conditional probabilities of whole
output traces given whole input traces. 
O'Neill et al.~\cite{ONeill:06:CSF}, based on Wittbold and
Johnson~\cite{Wittbold:90:SAP}, improve on batch-job models by
introducing strategies.
The strategy functions of O'Neill et al.~are deterministic, whereas
ours are probabilistic.

Clark and Hunt~\cite{ClarkH08}, following O'Neill et al., investigate
a hierarchy of strategies.
\emph{Stream strategies}, at the bottom of the hierarchy, are
equivalent to having agents provide all their inputs before system
execution as a stream of values.
But probabilities are essential for information-theoretic
quantification of information flow.
Clark and Hunt do not address quantification, instead focusing on the
more limited problem of noninterference.

The work of Shokri et al.
\cite{shokri11location} strives to quantify the privacy of users of
location-based services using Markov models and various machine
learning techniques for constructing and applying them.
Shokri et al.'s work employs two phases, one for learning a model of
how a principal's location could change over time, and one for
de-anonymizing subsequently observed, but obfuscated, location
information using this model.
Our work focuses on information theoretic characterizations of
security in such applications, and allows for the quantification
of how much information is learned about the strategies themselves.


\showindraft{
\newpage
}
\section{Conclusion}
\label{sec:conclusion}


In this paper we generalized the representation of the adversary's
prior knowledge about the secret from a single probability
distribution on secrets to an environment, which is a distribution 
on strategies for generating secrets.
This generalization allowed us to derive relevant extensions of the
traditional approaches to QIF, including measures of environmental
vulnerability, strategy vulnerability, and to disentangle 
\realsecurity and \aggrsecurity, two concepts usually conflated
in traditional approaches to QIF.

We are currently working on the extending the notion of strategies
to model secrets that evolve over time, and on the corresponding 
quantification of strategy leakage when secrets are processed
by a system.



\paragraph*{Acknowledgments}
 \begin{small}
   This work was developed with the support of CNPq, CAPES, FAPEMIG,
   US National Science Foundation grant CNS-1314857, and DARPA and the
   Air Force Research Laboratory, under agreement numbers
   FA8750-16-C-0022, FA8750-15-2-0104, and FA8750-15-2-0277. 
   The U.S. 
   Government is authorized to reproduce and distribute reprints for
   Governmental purposes not withstanding any copyright notation
   thereon. 
   The views, opinions, and/or findings expressed are those of the
   author(s) and should not be interpreted as representing the
   official views or policies of DARPA, the Air Force Research
   Laboratory, or the U.S. 
   Government.
 \end{small}

\pxm{20 pages max}

\showindraft{
\vfill\pagebreak
}

\printbibliography

\ifdefined\modereport

\appendix

\newpage
\section{Notes on hyper-distributions.}
\label{sec:hypers-properties}


Here we establish properties of hyper-distributions
used in our proofs of technical results.

Recalling our notation, let $\calx{=}\{x_{1}, x_{2}, \ldots, x_{n} \}$ 
be a set of secret values, and $\distsymb{\calx}$ be the set of
all probability distributions on $\calx$.
Let 
$\cals_{\calx}{=}\{ \strat_{1}, \strat_{2}, \ldots, \strat_{\m} \} \subset \distsymb{\calx}$ 
be a set of strategies of interest.%
\footnote{Given that $\calx$ is finite, we can make 
$\cals_{\calx}$ finite via a discretization that 
defines an indivisible amount $\mu$ of probability mass 
that strategies can allocate among secrets.
Any precision in strategies can be achieved 
by making $\mu$ as small as needed.}
We use letters like $\hyperh$, $\model$, $\env$ to 
denote hypers, and, unless stated otherwise, we will 
concentrate on hypers that are distributions on strategies
of interest, i.e., on hypers of type $\distsymb{\cals_{\calx}}$.%
\footnote{Note that $\distsymb{\cals_{\calx}} \subset \distsymb^{2}{\calx}$,
so in general all definitions and results that apply to objects of type
$\distsymb^{2}{\calx}$ will also apply to objects of type $\distsymb{\cals_{\calx}}$.}
We denote by $\strat_{j}(x_{i})$ the probability of secret 
$x_{i}$ being generated by strategy $\strat_{j}$, and 
by $\hyperh(\strat_{j})$ the (outer) probability of strategy 
$\strat_{j}$ being used in hyper $\hyperh$.


A hyper 
$\hyperh$
induces a joint 
probability distribution 
$p^{\hyperh}{:}\distsymb{(\calx{\times}\cals_{\calx})}$
on secrets and strategies, obtained as 
$p^{\hyperh}(x_{i},\strat_{j}){=}\hyperh(\strat_{j}) \strat_{j}(x_{i})$.
Conversely, given the joint distribution $p^{\hyperh}(x_{i},\strat_{j})$
we can recover the corresponding hyper $\hyperh$ using marginalization
and conditioning in the usual way.
More precisely, the outer distribution of hyper $\hyperh$
is recovered as $\hyperh(\strat_{j}){=}\sum_{x_{i}} p(x_{i},\strat_{j})$
for each strategy $\strat_{j}$,
and the probability of any secret $x_{i}$ given strategy $\strat_{j}$
is recovered as $\strat_{j}(x_{i}){=}\nicefrac{p^{\hyperh}(x_{i},\strat_{j})}{\hyperh(\strat_{j})}$
if $\hyperh(\strat_{j}){>}0$. 
Note that if $\hyperh(\strat_{j}){=}0$, we can 
pick any inner distribution $\strat_{j}{\in}\cals_{\calx}$ 
to be the corresponding inner without altering the hyper
(for instance, we can take the uniform inner in which 
$\strat_{j}(x_{i}){=}\nicefrac{1}{|\calx|}$ for every $x_{i}$).
For that reason, we will often equate a hyper $\hyperh$
with its joint distribution $p^{\hyperh}$.

Using marginalization and conditioning in the usual way,
a joint distribution $p^{\hyperh}{:}\distsymb{(\calx{\times}\cals_{\calx})}$
corresponding to a hyper can also be decomposed into 
a prior distribution $\prior_{\hyperh}$ on secrets and a conditional distribution 
$p^{\hyperh}_{S \mid X}$ of strategies given secrets.
More precisely, $\prior_{\hyperh}{=}\sum_{j} \hyperh(\strat_{j}) \strat_{j}$
is the prior consistent with hyper $\hyperh$,
and for each strategy $\strat_{j}$ and secret $x_{i}$
we have 
$p^{\hyperh}_{S \mid X}(\strat_{j}\mid x_{i}){=}
\nicefrac{p^{\hyperh}(x_{i},\strat_{j})}{\prior_{\hyperh}(x_{i})}$
if ${\prior_{\hyperh}(x_{i})}{>}0$.
Note that if $\prior_{\hyperh}(x_{i}){=}0$, we can 
pick any inner distribution on $\cals_{\calx}$ to be the
corresponding $\strat_{j}$ without altering the hyper
(for instance, we can make take the uniform inner such that 
$p^{\hyperh}_{S \mid X}(\strat_{j} \mid x_{i}){=} 
\nicefrac{1}{|\supp{\hyperh}|}$).

The decomposition of a hyper into a prior on secrets
and conditional probabilities on strategies given 
secrets can also be represented in a matrix form as follows.
Let $\joint{\hyperh}$ denote the $|\calx|{\times}|\cals_{\calx}|$
matrix corresponding to hyper $\hyperh$ in that 
$\joint{\hyperh}({i,j}){=}p^{\hyperh}(x_{i},\strat_{j})$.
Let $\dprior_{\hyperh}$ denote the $|\calx|{\times}|\calx|$ 
diagonal matrix in which, for all $1{\leq}i,j{\leq}|\calx|$,
$\dprior_{\hyperh}({i,i}){=}\prior_{\hyperh}(x_{i})$, and
$\dprior_{\hyperh}({i,j}){=}0$, whenever $i{\neq}j$.
Define $\Delta_{\hyperh}$ as a 
$|\calx|{\times} |\cals_{\calx}|$ matrix in which 
$\Delta_{\hyperh}({i,j}){=}p^{\hyperh}_{S \mid X}(\strat_{j} \mid x_{i})$.
Clearly $\Delta_{\hyperh}$ is a channel matrix mapping
secrets to strategies.
We can show that any hyper $\hyperh{:}\distsymb{\cals_{\calx}}$
can be obtained as the matrix multiplication
\begin{align}
\label{eq:hyper-multiplication}
\joint{\hyperh} = \dprior_{\hyperh} \cdot \Delta_{\hyperh},
\end{align}
since for all $x_{i}{\in}\calx$ and $\strat_{j}{\in}\cals_{\calx}$:
\begin{align*}
\joint{\hyperh}({i,j}) 
&= \sum_{k} \dprior_{\hyperh}({i,k}) \Delta_{\hyperh}({k,j}) & (\text{def. of matrix multiplication}) \\
&= \dprior_{\hyperh}({i,i}) \Delta_{\hyperh}({i,j}) & \text{($\dprior_{\hyperh}({i,k}) \neq 0$ only when $i=k$)} \\
&= \prior_{\hyperh}(x_{i}) \cdot p^{\hyperh}_{S \mid X}(\strat_{j} \mid x_{i}) & \text{(by def. of $\dprior_{\hyperh}$ and $\Delta_{\hyperh}$)} \\
&= p^{\hyperh}(x_{i},\strat_{j}).
\end{align*}

\begin{example}
The hyper $\hyperh$ below
\begin{align*}
\begin{array}{c||c|c|c}
& \strat_{1} & \strat_{2} & \strat_{3} \\ \hline \hline
x_{1}  & 1  & 0               & \nicefrac{1}{2}  \\
x_{2}  & 0  & 1            & \nicefrac{1}{2}  \\ \hline \hline
\hyperh & \nicefrac{1}{4} & \nicefrac{1}{4} & \nicefrac{1}{2}
\end{array}
\end{align*}
can be written as a joint matrix $\joint{\hyperh}$ and as the
product of a prior $\prior_{\hyperh}$ and a channel matrix 
$\Delta_{\hyperh}$ from secrets to strategies as follows.
\begin{align*}
\underbrace{
\mat{
\begin{array}{ccc}
\nicefrac{1}{4} & 0 & \nicefrac{1}{4} \\
0 & \nicefrac{1}{4} & \nicefrac{1}{4} \\
\end{array}
}
}_{\joint{\hyperh}}
=
\underbrace{
\mat{
\begin{array}{cc}
\nicefrac{1}{2} & 0 \\
0 & \nicefrac{1}{2} \\
\end{array}
}
}_{D{\prior_{\hyperh}}}
\cdot 
\underbrace{
\mat{
\begin{array}{ccc}
\nicefrac{1}{2} & 0 & \nicefrac{1}{2} \\
0 & \nicefrac{1}{2} & \nicefrac{1}{2} \\
\end{array}
}
}_{\Delta_{\hyperh}}
\end{align*}
\qed
\end{example}

Putting the above observations together, we see that
any hyper $\hyperh$ can be expressed as the result
of pushing a prior through a channel $\Delta_{\hyperh}$ as
follows.
\begin{restatable}{proposition}{reshyperdecomposition}
\label{prop:hyper-decomposition}
Every hyper $\hyperh{:}\distsymb{\cals_{\calx}}$ can be written as
\begin{align*}
\hyperh = [\prior_{\hyperh},\Delta_{\hyperh}],
\end{align*}
where $\prior_{\hyperh}{=}\E_{}\hyperh$ 
is the consistent prior with $\hyperh$,
and $\Delta_{\hyperh}$ is the $|\calx|{\times}|\cals_{\calx}|$ 
channel matrix from secrets to strategies in which 
$\Delta_{\hyperh}({i,j}){=}p^{\hyperh}_{S \mid X}(\strat_{j} \mid x_{i})$.
\end{restatable}

A direct consequence of Proposition~\ref{prop:hyper-decomposition},
is that the environmental vulnerability of a hyper $\hyperh$ can be 
expressed as the posterior $g$-vulnerability of channel 
$\Delta_{\hyperh}$ for hyper $\prior_{\hyperh}$.

\begin{restatable}{corollary}{reshypervg}
\label{cor:hypervg}
For any vulnerability $\priorvf$ and any hyper $\hyperh{:}\distsymb{\cals_{\calx}}$,
there exists a $\vgf$ such that
\begin{align*}
\contextv{\hyperh} 
&= \vgh{\prior_{\hyperh},\Delta_{\hyperh}}.
\end{align*}
\end{restatable}

The next result shows that if a $\hyperh'$ is an abstraction 
of another hyper $\hyperh$ via aggregation matrix $A$,
then the channel $\Delta_{\hyperh'}$ from secrets to strategies
in $\hyperh'$ can be obtained as the cascading of the channel 
$\Delta_{\hyperh}$ from secrets to strategies in $\hyperh$
and the aggregation matrix $A$.

\begin{restatable}{lemma}{resaggregationDelta}
\label{lemma:aggregation-Delta}
For any aggregation matrix $A$, 
$\hyperh'{=}\hyperh A$  iff
$\Delta_{\hyperh'}{=}\Delta_{\hyperh} A$.
\end{restatable}

\begin{proof}
Assume $\hyperh'{=}\hyperh A$ for some aggregation matrix $A$.
Then, by Proposition~\ref{prop:aggregation} we know 
that $\prior_{\hyperh'}{=}\prior_{\hyperh}$.
Hence 
\begin{align*}
\hyperh
&= \hyperh' A & \text{(by hypothesis)} \nonumber \\
&= D\prior_{\hyperh'} \Delta_{\hyperh'} A & \text{($\hyperh'{=}D\prior_{\hyperh'} \Delta_{\hyperh'}$ by Equation~\eqref{eq:hyper-multiplication})} \nonumber \\
&= D\prior_{\hyperh} \Delta_{\hyperh'} A & \text{(by Proposition~\ref{prop:aggregation})}. 
\end{align*}

But notice that we can decompose $\hyperh$ into $D\prior_{\hyperh} \Delta_{\hyperh}$.
Combining this 
with the derivation above,
we conclude that
$\Delta_{\hyperh}{=}\Delta_{\hyperh'} A$.
\qed
\end{proof}

An important observation is that the construction of 
models via aggregation matrices means that the adversary's 
identification of a strategy as a context of execution is 
independent from the actual secret value, given the actual 
strategy.
Formally, if $\model{\abstracts}\env$ then for all secrets $x$,
strategies $ \m$ in $\model$ and strategies $\strat$ in $\env$:
$
p(\m\mid \strat, x){=}p(\m\mid \strat)
$.
As a direct consequence, the probability of any secret is 
independent of the abstracted model, given the original 
model.

\begin{restatable}{lemma}{resprobs}
  \label{lemma:probs} \pxm{Perhaps this can be called
    \emph{consistency}?}
  If $\model{\abstracts}\env$ then for all secrets $x$, strategies $ \m$
  in $\model$, and strategies $\strat$ in $\env$,
$
p(x \mid  \m, \strat){=}p(x \mid \strat){=}\strat(x)
$.
\end{restatable}

\begin{proof}
If $\model{\abstracts}\env$ then for all secrets $x$,
strategies $ \m$ in $\model$ and strategies $\strat$ 
in $\env$ we have
$
p(\m\mid \strat, x){=}p(\m\mid \strat)
$.
Hence, also for all $x$, $\m$, $\strat$:
\begin{align*}
p(x \mid \m, \strat)
&= \frac{p(x,\m,\strat)}{p(\m,\strat)} & \text{(by definition of conditional prob.)} \\
&= \frac{\env(\strat) \cdot \strat({x}) \cdot p(\m \mid x, \strat)}{\env(\strat) \cdot p(\m \mid \strat)} 
& \text{(by the chain rule for probabilities)} \\
&= \strat_{x} & \text{(since $p(\m \mid x, \strat) = p(\m \mid \strat)$)} \\
&= p(x \mid \strat).
\end{align*}
\qed
\end{proof}

\showindraft{
\newpage
}
\section{Proofs of technical results.}
\label{sec:proofs}


\subsection{Preliminaries for proofs}

The following definitions and results from the literature will be needed in 
our proofs.

\begin{theorem}[Jensen's inequality]
\label{theo:jensen}
If $f$ is a convex ($\smile$) function,
$\lambda_{1}, \lambda_{2}, \ldots, \allowbreak \lambda_{n}$ are convex coefficients, and
$x_{1}, x_{2}, \ldots, x_{n}$ are arbitrary, then
\begin{align*}
f\left(\sum_{i} \lambda_{i} x_{i}\right)
&\leq \sum_{i}\lambda_{i} f(x_{i}).
\end{align*}
In particular, if $X$ is a random variable and $f$ is a convex
($\smile$) function,
\begin{align*}
f\left(\E{X}\right) 
&\leq \E f(X).
\end{align*}
\end{theorem}

\begin{theorem}[\qm{Miracle}~\cite{alvim12gain}]
\label{theo:miracle}
For any channel $C$ and $g$-vulnerability $\vgf$,
\begin{align*}
\max_{\prior} \left( \log_{2}{\frac{\vgf[\prior,C]}{\vg{\prior}}} \right)
&\leq \log_{2}{\frac{V^{(Bayes)}[\prior,C]}{V^{(Bayes)}{(\prior})}}.
\end{align*}
\end{theorem}

The next result states that $g$-vulnerabilities 
provide a precise 
information-theoretic meaning to the 
\emph{composition refinement relation} on channels.
We say a channel $C$ from $\calx$ to $\caly$ 
\emph{(composition) refines a channel} $C'$ from 
$\calx$ to $\calz$ (or, equivalently, that $C'$ 
\emph{is (composition) refined by} $C$), 
denoted by $C'{\compref}C$, if there is a channel matrix 
$R$ from $\caly$ to $\calz$ such that $C'{=}CR$.
The next result states that a channel $C$ 
refines a channel $C'$ iff for all gain functions
and priors, the posterior $g$-vulnerability of the
secret given channel $C'$ never exceeds that of
the secret given channel $C$.

\begin{theorem}[Composition refinement and 
strong $g$-vulnerability ordering 
\cite{alvim12gain,mciver14abstract}]
\label{theo:refinement-gorder}
For any two channel matrices $C$, $C'$,
$C'{\compref}C$ iff 
$\vgh{\prior,C'}{\leq}\allowbreak\vgh{\prior,C}$
for all priors $\prior$ and gain functions $g$.
\end{theorem}

\subsection{Proofs of Section~\ref{sec:strategies}}

\rescollapse*

\begin{proof}
For all point-hyper environments $\env$:
\begin{align*}
\contextv{\env} 
&= \E_{\env} \priorvf & \text{(by definition of $\contextvf$)} \\
&= 1 \cdot \priorv{\prior} & \text{(since $\env=[\prior]$)} \\
&= \priorv{\prior}.
\end{align*}
\qed
\end{proof}

\subsection{Proofs of Section~\ref{sec:security-real-vs-aggregation}}

\resenvironmentvsdistv*
\begin{proof}
Given that $\contextvf$ is a convex function, for every
environment $\env$ we can make use of Jensen's inequality as follows.
\begin{align*}
\contextv{\env}
&= \E_{\env} \priorvf & \text{(by definition of $\contextvf$)} \\
&\geq \priorvf \left(\E{\env}\right)  & \text{(by Jensen's inequality and convexity of $\contextvf$)} \\
&= \priorvf \left(\prior_{\env}\right)  & \text{(by definition of $\prior_{\env}$)}
\end{align*}
\qed
\end{proof}

\resstrvlowerbound*
\begin{proof}
By Corollary~\ref{cor:hypervg}, for any environment $\env$ 
$\strv{\env}$ can be written as
\begin{align*}
\strv{\env}
&= \frac{\vg{\prior_{\env}}}{\vgh{\prior_{\env},\Delta_{\env}}} \\
&\geq \frac{V^{(Bayes)}(\prior_{\env})}{V^{(Bayes)}{[\prior_{\env},\Delta_{\env}]}}
& \text{(by Theorem~\ref{theo:miracle}, since $\log_{2}$ is monotonic)} \\
&= \frac{\bayesv{\prior_{\env}}}{\contextbayesv{\env}} & \text{(by Corollary~\ref{cor:hypervg})}
\end{align*}
\qed
\end{proof}

\subsection{Proofs of Section~\ref{sec:models}}

\resaggregation*

\begin{proof}
By Proposition~\ref{prop:hyper-decomposition}, $\hyperh$
can be decomposed as $[\prior_{\hyperh},\Delta_{\hyperh}]$.
If $\hyperh'{=}\hyperh{\cdot}A$, then by Lemma~\ref{lemma:aggregation-Delta}
it can be expressed as $\hyperh'{=}[\prior_{\hyperh},\Delta_{\hyperh}{\cdot}A]$,
from which the result follows.
%
\qed
\end{proof}

\resenvironmentgen*

\begin{proof}
First, recall that by Lemma~\ref{lemma:probs}, 
in any abstraction $\model$ for environment $\env$,
for all $x$, $\m$, $\strat$ we have $p(x \mid \m, \strat){=}p(x \mid \strat){=}\strat(x)$,
and, hence the joint distribution is given by
\begin{align}
p(\strat,\m,x) 
&= p(\strat) p(\m \mid \strat) p(x \mid \m, \strat) & \text{(by the chain rule for prob.)} \nonumber \\
&= p(\strat) p(\m \mid \strat) p(x \mid \strat) & \text{(by the observation above)} \nonumber \\
&= \env(\strat) \cdot A({\strat,\m}) \cdot \strat({x}). \label{eq:prop-environment-gen1}
\end{align}

Hence we can derive
\begin{align*}
\modelv{\model}{\env}
&= \sum_{\strat} \env(\strat) \sum _{ \m} A({\m,\strat}) \sum_{x} \strat({x}) \, g(w_{ \m},x)
& \text{(definition of $\modelvf$)}\\
&= \sum_{\strat,\m,x} p(\strat,\m,x) g(w_{\m},x) & \text{(by Equation~\eqref{eq:prop-environment-gen1})} \\
&= \sum_{\m} p(\m) \sum_{x} p(x \mid \m) \sum_{\strat} p(\strat \mid \m,x) g(w_{\m},x) & \text{(chain rule of prob.)}\\
&= \sum_{\m} \model(\m) \sum_{x} \m(x) g(w_{\m},x) \sum_{\strat} p(\strat \mid \m,x) \\
&= \sum_{\m} \model(\m) \sum_{x} \m(x) g(w_{\m},x) & \text{($\textstyle \sum_{\strat} p(\strat \mid \m,x) = 1$)}\\
&= \vgf(X \mid \model) \\
&= \contextv{\model}
\end{align*}
\qed
\end{proof}

The following auxiliary result states that the more refined
is an abstraction for an environment, the greater the 
corresponding environmental vulnerability.

\begin{restatable}{lemma}{resgorder}
\label{lemma:g-order}
If $\model', \model$ are abstractions for an environment $\env$,
then $\model'{\abstracts}\model$ iff
$\contextv{\model'}{\leq}\contextv{\model}$ for all $\priorvf$.
\end{restatable}

\begin{proof}
If $\model'{\abstracts}\model$, by Proposition~\ref{prop:hyper-decomposition}
and Lemma~\ref{lemma:aggregation-Delta} we know there is 
a prior $\prior$ and channel matrices $\Delta_{\model}$ and $A$ such that
$\model{=}[\prior_{M},\Delta_{\model}]$ and
$\model'{=}[\prior_{M},\Delta_{\model}A]$.
By Theorem~\ref{theo:refinement-gorder} we know that 
$\vgf[\prior_{\model},\Delta_{\model} A]{\leq}\vgf[\prior_{\model'},\Delta_{\model'}]$,
for all $\vgf$, and hence by Corollary~\ref{cor:hypervg} we have that
$\contextv{\model'}{\leq}\contextv{\model}$.
\qed
\end{proof}

\resordermodels*

\begin{proof}
Just combine Lemma~\ref{lemma:g-order} with 
Proposition~\ref{prop:environmentv-gen}.
\qed
\end{proof}

\resstrategyvulnraiblityrefinement*

\begin{proof}
From Lemma~\ref{lemma:g-order} we know that
$\model'{\abstracts}\model$ iff 
$\contextv{\model'}{\leq}\contextv{\model}$
for all $\priorvf$.
By noting that 
$\strvmdl{\model'}{\env}{=}\nicefrac{\contextv{\model'}}{\contextv{\env}}$ and
$\strvmdl{\model}{\env}{=}\nicefrac{\contextv{\model}}{\contextv{\env}}$, 
and $\contextv{\env}$ is the same in non-negative value in the denominator
of both fractions, the result follows.
\qed
\end{proof}

\resboundsstrvgivenmodel*

\begin{proof}
Note that the most refined abstraction of any
environment $\env$ is the environment itself,
and that the least refined abstraction is the
point-hyper $[\prior_{\env}]$.
This implies, from Proposition~\ref{prop:strategy-vulnerability-refinement},
that the lower bound on $\strvmdl{\env}{\model}$ is achieved when 
$\model{=}[\prior_{\env}]$, so
\begin{align*}
\strvmdl{\env}{[\prior_{\env}]}
&= \frac{\contextv{[\prior_{\env}]}}{\contextv{\env}} & \text{(by definition of $\strvf$ given an abstraction)} \\
&= \frac{\priorv{\prior_{\env}}}{\contextv{\env}} & \text{(by Proposition~\ref{prop:collapse})} \\
&= \strv{\env}, & \text{(by definition of $\strvf$)}
\end{align*}
and the upper bound on $\strvmdl{\env}{\model}$ is
achieved when $\model{=}\env$, so
\begin{align*}
\strvmdl{\env}{\env}
&= \frac{\contextv{\env}}{\contextv{\env}} & \text{(by definition of $\strvf$ given an abstraction)} \\
&= 1.
\end{align*}
\qed
\end{proof}

\subsection{Proofs of Section~\ref{sec:hypers-expressiveness}}

In this section we need the following definitions and notation.
A \emph{hyper of degree $n$}, or \emph{$n$-hyper}, 
on $\calx$ is an object of type $\distsymb^{n}\calx$, 
where $\distsymb^{1}\calx{=}\distsymb{\calx}$ and 
$\distsymb^{n}\calx{=}\allowbreak\distsymb\left(\distsymb^{n-1}\calx\right)$ 
for $n \geq 2$.
We let $\hypn{n}$ range of hypers of degree $n$.
In particular, $\hypn{1}{:}\distsymb{\calx}$ is a prior 
on secrets, and $\hypn{2}{:}\distsymb^{2}{\calx}$ is a 
hyper-distribution.
Any $n$-hyper $\hypn{n}$ can be represented as a joint 
distribution $\jointh{\hypn{n}}{X{\times}{Y_{1}}{\times}\ldots{\times}Y_{n-1}}$
on the Cartesian product 
$\calx{\times}\caly_{1}{\times}\caly_{2}{\times}\ldots{\times}\caly_{n-1}$
in which each $\caly_{i}$ is a set of labels for all hypers $\hypn{i{-}1}$ 
of order $i{-}1$, and $Y_{i}$ is the associated random variable.
We may denote $\jointh{\hypn{n}}{X{\times}{Y_{1}}{\times}\ldots{\times}Y_{n-1}}$
simply by $\jointh{\hypn{n}}{X{\times}Y}$.
Given a joint $\jointh{\hypn{n}}{X{\times}Y}$,
we denote by $\jointh{\hypn{n}}{X{\times}Y \mid  {}_{\geq}Y_{i}}$, for $1{\leq}i{\leq}n-1$, the joint 
$\jointh{\hypn{n}}{X{\times}{Y_{1}}{\times}\ldots{\times}Y_{i-1}\mid Y_{i}{\ldots}Y_{n-1}}$
on $\calx{\times}\caly_{1}{\times}\caly_{2}{\times}\ldots{\times}\caly_{i-1}$
obtained by marginalization of 
$\jointh{\hypn{n}}{X{\times}{Y_{1}}{\times}\ldots{\times}Y_{n-1}}$ w.r.t. 
$Y_{i}{\times}Y_{i+1}{\times}{\ldots}{\times}Y_{n-1}$.
Analogously, given a hyper $\hypn{n}$ of type ${\distsymb^{n}{\calx}}$,
for $1{\leq}i{\leq}n-1$ we denote by $\collapseh{i}$ the hyper of type 
${\distsymb^{i}{\calx}}$ corresponding to the marginalization of 
$\hypn{n}$ w.r.t. to $Y_{i}{\times}Y_{i+1}{\times}{\ldots}{\times}Y_{n-1}$.
Recall that we defined a \emph{vulnerability of order $n$} as 
a function $\hypnvf{n}{:}\distsymb^{n}\calx{\rightarrow}\reals$ 
s.t. $\hypnv{1}{\hypn{1}}{=}\priorv{\hypn{1}}$, and
$\hypnv{n}{\hypn{n}}{=}\E_{\hypn{n}}\hypnvf{n-1}$ for $n \geq 2$.
In particular, $\hypnv{1}{\hypn{1}}{=}\allowbreak\priorv{\hypn{1}}$ is the 
traditional vulnerability on secrets, 
and $\hypnv{2}{\hypn{2}}{=}\E_{\hypn{2}}\priorvf$ 
is environmental vulnerability.

The next result shows that an adversary who reasons according 
to a model of type $\distsymb^{n}{\calx}$ for some $n{\geq}2$
can do as great as a job as an adversary with an appropriate
model of type $\distsymb^{2}{\calx}$.

\reshypersexpressivenes*

\begin{proof}
By induction on the degree $n$ of the hyper.
For the base case, note that $\hypn{2}{=}\collapseh{2}$ and,
hence, $\hypnv{2}{\hypn{2}}{=}\hypnv{2}{\collapseh{2}}$.
As for the inductive case, for any $n{\geq}2$, 
$\hypnv{n}{\hypn{n}}$ can be written as:
\begin{align*}
&  \E_{\hypn{n}}\hypnvf{n-1} \\
&= \text{(expanding the summation)} \\
&  \sum_{y_{n-1}{\in}\caly_{n-1}} 
   \jointh{\hypn{n}}{X{\times}Y}(y_{n-1}) 
   \hypnv{n-1}{\jointh{\hypn{n}}{X{\times}Y \mid  {}_{\geq}Y_{n-1}}} \\
&= \text{(by definition of $\hypnvf{n-1}$)} \\
&  \sum_{y_{n-1}{\in}\caly_{n-1}} 
   \jointh{\hypn{n}}{X{\times}Y}(y_{n-1}) 
   \sum_{y_{n-2}{\in}\caly_{n-2}} 
   \jointh{\hypn{n}}{X{\times}Y\mid  {}_{\geq}Y_{n-1}}(y_{n-2}) 
   \hypnv{n-2}{\jointh{\hypn{n}}{X{\times}Y\mid  {}_{\geq}Y_{n-2}}} \\	
&= \text{(groupping the summation)}\\
&  \sum_{(y_{n-1},y_{n-2}){\in}\caly_{n-1}{\times}{\caly_{n-2}}} 
   \jointh{\hypn{n}}{X{\times}Y \mid _{\geq}Y_{n-2}}(y_{n-1},y_{n-2}) 
   \hypnv{n-2}{\jointh{\hypn{n}}{X{\times}Y \mid _{}{\geq}Y_{n-2}}} \\ 
&= \text{(by definition of $\collapseh{n-1}$)} \\
&  \hypnv{n-1}{\collapseh{n-1}} \\
&= \text{(by the induction hypothesis)} \\
&  \hypnv{2}{\collapseh{2}}.
\end{align*}
\qed
\end{proof}

\showindraft{
\newpage
\section{Reviews from POST}
\label{sec:reviews-post}

\input{nonfinaltex/reviews-post}

%

\newpage
\section{Notes on directions for future work}
\label{sec:sargasso}

\input{nonfinaltex/appendix-sargasso}
}

%
\fi

\end{document}